 \documentclass[preprint,aps,showpacs,nofootinbib]{revtex4}
\usepackage{graphicx,amsmath,amssymb}
\usepackage{psfrag}
\usepackage{color}
\usepackage{hyperref}

\hypersetup{
	colorlinks=true,
	linkcolor=black,
	citecolor=red,			
	urlcolor=blue,			
	filecolor=blue}
\newcommand{\be}{\begin {eqnarray}}
\newcommand{\ee}{\end{eqnarray}}
\usepackage[normalem]{ulem}
\usepackage[T1]{fontenc}
\usepackage[utf8]{inputenc}
\usepackage{lmodern}
\usepackage{amsfonts}
\newcommand{\BC}{Borici-Creutz}
\begin{document}


\title{Mixed action with {\BC} fermions on staggered sea}
\author{S. Basak$^1$, D. Chakrabarti$^2$, J. Goswami$^2$}
\affiliation{
$^1$School of Physical Sciences, NISER Bhubaneswar, Khurda-752050, India.\\
$^2$Department of Physics, Indian Institute of Technology Kanpur,
Kanpur-208016, India}
\date{\today}
\begin{abstract}
Mixed action lattice QCD with {\BC} valence quarks on staggered sea is
investigated. The counter terms in {\BC} action are fixed nonperturbatively
to restore the broken parity and time symmetries.
When symmetries are restored, the usual signatures of partial
	quenching are recovered. We find the scalar correlators to be
	negative for lower valence quark masses but the errorbars are
	rather large when their mean values are negative at earlier
	time slices. The size of unitarity violation
due to different discretization of valence and sea quark is determined
by measuring $\Delta_{\rm mix}$ and is found to be comparable with other 
mixed action studies.
\end{abstract}

\pacs{11.15.Ha, 11.30.Rd}

\maketitle

\section{Introduction}
Minimally doubled fermions are promising techniques to study light
quarks on lattice. They are known to preserve chiral symmetry for a
degenerate quark doublet and are local. This can be helpful for $N_f
=2$ lattice simulations, and is relatively simpler and possibly faster
than Ginsparg-Wilson fermions. There are two main realizations of the
minimally doubled fermions -- Karsten-Wilczek \cite{K,W} and Borici-Creutz 
\cite{creutz,borici}. In this paper, we study mixed action lattice QCD
with Borici-Cruetz valence quarks in Asqtad improved staggered sea,
restricting to light quarks only. 

Motivated by the fact that electrons on a graphene lattice are described
by a massless Dirac-like equation (quasi-relativistic Dirac equation),
Creutz  proposed a four dimensional Euclidean lattice action describing
two flavors of fermion, centered at $\pm p_\mu$ in the momentum space
\cite{creutz}. The action was defined on a honeycomb or graphene lattice
with tunable parameters to control the magnitude of $p_\mu$. Borici
found a solution for the  parameters such that the two flavors are
located at $p=(0,0,0,0)$ and $(\pi/2,\pi/2,\pi/2,\pi/2)$. The chirally
invariant Borici-Creutz action breaks hypercubic symmetry leading to
the breaking of discrete symmetries like parity and time-reversal
\cite{bedaque,creutz2}. This introduces non-covariant counterterms
through quantum corrections. The renormalization properties
of the BC fermion at  one loop in perturbation theory have been
investigated in \cite{capitani1,capitani}. It was shown that in presence
of gauge background with integer-valued topological charge, BC action
satisfies the Atiyah-Singer index theorem \cite{dc}. The 
cut-off effects on BC fermions at the tree-level of perturbation theory has been 
studied by K. Cichy et. al.\cite{cichy}.

However, there is a dearth of numerical studies with Borici-Creutz
fermions and very few literatures which would suggest its 
usefulness
and performance in lattice QCD simulations. In recent times, we have
initiated a series of  detailed numerical studies
to ascertain its viability in lattice simulations. Using Borici-Creutz
fermions we have studied discrete chiral symmetry breaking in a two
dimensional Gross-Neveu model \cite{GCB1} and the
mass spectroscopy in 2-dimensional field theories \cite{GCB2} in
lattice. In this work, we extend our investigations to 4-dimensional
lattice and  attempt
to simulate QCD with Borici-Creutz action. Here, we consider a mixed
action approach with Borici-Creutz valence quarks
on an Asqtad improved staggered sea. The coefficients of the counterterms
of the renormalized action are fixed non-perturbatively to restore the
broken hypercubic symmetry of the Borici-Creutz action.

At finite lattice spacing, mixed action QCD violates unitarity but it
is believed to have the correct continuum limit. In mixed action, proper
matching of the sea and valance actions is important. One can  tune the
valence quark masses to have the desired meson masses in the mixed
action to agree with QCD with unitarity. Since the
valence and sea quarks have different discretization effects,
it is not, however, possible to tune the mesons made up of purely
valence and sea quarks to restore unitarity at finite lattice spacing.
Alternatively, one can correct the mismatch in the bare parameters
using partially quenched mixed action approach in which the actions
are not matched but the parameters are tuned to reproduce the physical
observables\cite{TM}.

The scalar correlator is known to be sensitive to the unitarity
violation, due to different discretization of valence and sea quarks,
because of two mesons intermediate states in the hair-pin diagram.
After the counterterms are tuned and the breaking of the hypercubic
symmetry minimized, the scalar correlator shows
the effect of partially quenched QCD and becomes negative for valance
quark mass smaller than sea quark mass ($m_{\rm val} < m_{\rm sea}$).
The quark mass dependence of pion mass squared is
found to be linear for heavier quarks but logarithmic corrections
as predicted by one loop chiral perturbation theory become prominent
for lighter quarks. The lattice result for the pion
mass agrees well with the one loop partially quenched
chiral perturbation prediction.
 There are several studies of mixed action lattice QCD with different
combinations of sea and valence quarks \cite{DWF1,DWF2, Clover, Overlap1,
Overlap2,Mixed1,Mixed2,Mixed3,mixed4, TM,twisted_mass}. Many of these studies 
use Asqtad
staggered sea quarks.

In the leading order mixed action chiral perturbation theory, a lattice
spacing dependent low energy constant $\Delta_{\rm mix}$ appears as a
free parameter in the mass formula for meson made up of one valence and
one sea quark. The $\Delta_{\rm mix}$ gives a measure of the unitarity
violation. (In this paper we actually calculate $\tilde{\Delta}_{\rm mix}$
which is different from $\Delta_{\rm mix}$ by an additional term as will
be explained later in the text in Sec,\ref{sec_deltamix}.) The $\Delta_{\rm mix}$ obtained in this work
is found to be comparable to the mixed action with domain wall fermions
on staggered sea \cite{DWF1,DWF2}. Fermions based on the solutions of
Ginsparg-Wilson relation, such as Domain-wall or Overlap fermions, are
in general computationally very demanding but Borici-Creutz fermions
being ultra local is expected to be computationally cheaper. So, it
raises the hope that Borici-Creutz fermion might be a good alternative
for QCD simulations with dynamical fermions.




\section{Borici-Creutz action and Point split method}

\noindent
The free {\BC} action in discretized 4 dimensional space-time lattice
is written as,
\be
S_{BC}&=& \sum_{x} \left[ \frac{1}{2}\sum_{\mu}\bar{\psi}(x)\gamma_\mu
(\psi(x+\hat{\mu})-\psi(x-\hat{\mu})) - \right. \nonumber \\
&& \left. \frac{i}{2}\sum_{\mu}\bar{\psi}(x)(\Gamma-\gamma_{\mu})
\left( 2\psi(x) - \psi(x+\hat{\mu})-\psi(x-\hat{\mu}) \right) +m
\bar{\psi}(x)\psi(x) \right] \label{BC}
\ee
where, $\Gamma=\frac{1}{2}(\gamma_{1}+\gamma_{2}+\gamma_{3}+\gamma_{4})$,
$\{\Gamma,\gamma_{\mu}\} =1$ and we have taken lattice spacing $a=1$.
In the momentum space, the action turns out to be diagonal and is,
\be
S_{BC}&=& \int \frac{d^4p}{(2\pi)^4} \bar{\psi}(p) \left[ \sum_\mu
\left( \gamma_\mu\sin(p_\mu)+i(\Gamma-\gamma^\mu)\cos(p_\mu) \right)
- 2i\Gamma + m \right] \psi(p). \label{bcmom}
\ee
The zeroes of the free massless Dirac operator are at $(0,0,0,0)$ and
$(\pi/2, \pi/2, \pi/2, \pi/2)$. The spinor $\psi(p)$ contains two degenerate
flavors. We can construct the fields for these two flavors at two different
poles by the method of point splitting \cite{creutz3, tiburzi,Zeqirllari}. We define 
\be
d(p) & \equiv & \frac{1}{4} \Gamma\sum_\mu (1-\sin{p_\mu})\psi(p)
\nonumber \\
u(p) & \equiv & \frac{1}{4} \sum_\mu (1-\cos(p_\mu+\pi/2))\psi(p+\pi/2)
\label{ud_sum} \\
\Rightarrow \;\; u(p-\pi/2) &=& \frac{1}{4} \sum_\mu (1-\cos p_\mu)
\psi(p),
\ee
such that $d(0)=\Gamma \psi(0)$ and $d(\pi/2)=0$ implying $d$ is defined
as an excitation around the pole at $(0,0,0,0)$. Similarly, $u(0)=\psi(
\pi/2)$ is defined around the pole at $(\pi/2, \pi/2, \pi/2, \pi/2)$.
The flavors are defined within an energy region, for $d$ quark we define
a region in the momentum space such that the energy $E < \pi/4$ and for
the $u$ quark $E>\pi/4$, so that there is no overlap between these two
fields. The definition of these fields are, however, not unique. One can
also choose a different prescription, 
\be 
d(p)& \equiv &\Gamma~\prod_\mu (1-\sin{p_\mu})\psi(p) \nonumber \\
u(p-\pi/2) & \equiv &  \prod_\mu (1-\cos p_\mu)\psi(p).
\label{ud_prod}
\ee
These two set of definitions of \autoref{ud_sum} and \autoref{ud_prod} are
equivalent to each other, but we found that the spectrum obtained
using the product form of fields in \autoref{ud_prod} to be noisier
compared to those obtained from \autoref{ud_sum}. Therefore, in our
work we have implemented the point split fields as defined in
\autoref{ud_sum}, where the summation over the Lorentz index $\mu$ is
carried out. The $\Gamma$ factor in $d$-field is inserted since the
chiral symmetry is flavored {\em i.e.}, $u$ and $d$ have a relative
minus sign under $\gamma_5$ transformation.

The Borici-Creutz action \autoref{BC} has a special direction in
Euclidean space which is the major hypercube diagonal (the line
joining the two zeros) given by $\Gamma$. The action is symmetric
under the cubic subgroup of the hypercubic group which preserves
the special direction. This breaks the reflection symmetry of the
hypercube leading to the breaking of parity and time symmetry\cite{bedaque}.
Because of the broken hypercubic symmetry, the counterterms are
necessary for a renormalized theory. The allowed counterterms for
the Borici-Creutz action are dimension-4 counterterm $c_4(g_0) \,
\bar{\psi} \Gamma \sum_\mu D_\mu \psi $ and dimension-3 counterterm
$ic_3(g_0) \bar{\psi} \Gamma \psi$ (for a discussion of counterterms
in the context of minimally doubled fermions see \cite{capitani}). Then the
complete renormalized action, with the gauge interaction turned on,
reads
\begin{eqnarray}
S_{BC} & = & \sum_{x} \left[ \frac{1}{2} \sum_\mu \left( \overline{\psi}
(x) \,(\gamma_\mu+c_4(g_0)\Gamma + i\gamma_\mu^\prime)\,U_\mu(x) \psi(x+
\hat{\mu}) \right. \right. \nonumber\\
 && \left. - \;\overline{\psi}(x+\hat{\mu}) \,(\gamma_\mu-c_4(g_0)\Gamma
- i\gamma_\mu^\prime) \,U_\mu^\dagger(x) \psi(x) \right) \nonumber\\
 && \left. + \; \overline{\psi}(x) \left( m + i\tilde{c}_3(g_0)\Gamma
\right) \,\psi(x) \right], \label{bc_full}
\end{eqnarray}
where, $\tilde{c}_3=c_{3}-2$, $\gamma^\prime_\mu=\Gamma-\gamma_\mu =
\Gamma\gamma_\mu\Gamma$ and $g_0$ is the bare coupling parameter. The
coefficients of the dimension-3 and 4 counterterms, $c_3$ and $c_4$
respectively, have been evaluated in 1-loop lattice perturbation theory
\cite{capitani} and are given by,
\begin{eqnarray}
c_3(g_0) & = & 29.54170 \cdot \frac{g_0^2}{16\pi^2}C_F+ \mathcal{O}
(g_0^4), \nonumber\\
c_4(g_0) & = & 1.52766 \cdot \frac{g_0^2}{16\pi^2}C_{F}+ \mathcal{O}
(g_0^4).\label{C_pt}
\end{eqnarray}
The mixed action study has been carried out with the above renormalized
Borici-Creutz action( \autoref{bc_full}) for the valence quarks.

\subsection{Simulation details}
\noindent
The mixed action simulation is carried out with Borici-Creutz fermions
on three ensembles of publicly available MILC lattices with 2+1 dynamical
flavors of Asqtad improved staggered fermions \cite{MILC}, with a fixed
ratio $am_l / am_s = 1/5$. The details of the MILC configurations used
in this work are listed in Table \ref{MLAT}.
\begin{table}[hpt] 
\centering \begin{tabular}{|c|c|c|c|c|c|}
\hline Lattice dim. & $\beta = 10/g_0^2$ & $am_l / am_s$ & $a$ (fm) &
volume (fm)$^3$ & \# configs \\ \hline
$16^3\times 48$ & 6.572 & 0.0097 / 0.0484 & $\approx$ 0.15 & $\sim$ (2.4
fm)$^3$ & 40 \\ \hline
$20^3\times 64$ & 6.76 & 0.01 / 0.05 & $\approx$  0.13  & $\sim$ (2.6
fm)$^3$ & 40 \\ \hline
$28^3\times 96$ & 7.09 & 0.0062 / 0.031 & $\approx$ 0.09  & $\sim$ (2.5
fm)$^3$ &30 \\ \hline
\end{tabular}
\caption{Details of MILC lattices \cite{MILC} used in this work.}
\label{MLAT}
\end{table}
We generated the staggered sea quark propagators with sea quark mass
$m_l$ and $m_s$ which are required for construction of the mixed
valence-sea mesons. The
number of configurations used are determined by the limitations of our
computational resources and by the reasonable size of statistical errors
of the meson masses. Since the measurement of $a^2\,\Delta_{\rm mix}$ is
one of the important parts of this study, we prefer to use lattices of
three different spacings in lieu of multiple $am_l/am_s$ ratios (for a
fixed lattice spacing) as it has been shown in multiple studies that
$\Delta_{\rm mix}$ does not depend significantly on the sea masses
\cite{Overlap2}.

The Borici-Creutz valence quark propagators are constructed using a
range of bare quark masses [0.0075 -- 0.5] which restricts the $m_\pi
L$ around 4 and we think best suited to study the chiral logs and
partial quenching.
The strange mass is tuned by setting the fictitious
$s\bar{s}$ pseudoscalar mass to 682 MeV \cite{Davies}. Similar strange mass is also obtained by  using $m_{\rm ps}
/ m_{\rm vec} = 0.673$ \cite{MILC} for tuning. 
Random wall sources are used for the generation of all quark propagators.
Both the APE and HYP smearing of the MILC lattices are tried but no
significant advantages have been observed. The meson masses are extracted
by fitting the meson propagators with double exponential ansatz. All the
fits are uncorrelated and the errors are from Jack-knife analysis.

\section{Non-perturbative fixing of the counter terms}
\subsection{Parity condensate}
The Borici-Creutz fermions exhibit parity-flavor breaking, which follows
from absence of the hypercubic symmetry of the action in \autoref{BC}. As
because the CPT is conserved, the $T$-symmetry is also broken. Consequently
the counterterms are necessary for renormalized theory and those that are
allowed by the remaining symmetry are added to the action. In the present
case, as discussed before, two such counterterms of dimension-3 and 4 are
introduced. The coefficients of these operators, $\tilde{c}_3(g_0)$ and
$c_4(g_0)$, can then be tuned to restore the desired symmetries. These
coefficients depend on the gauge coupling, the expressions for which in
1-loop lattice perturbation theory are given in \autoref{C_pt}. This means
these values will be different for different lattices. The perturbative
values of the coefficients corresponding to the three lattices that we
use in this work, are given in the  \autoref{pert}.
\begin{table}[h]
\centering
\begin{tabular}{|c|c|c|c|}
\hline 
Lattice size & $16^3\times 48$ & $20^3\times 64$ & $28^3\times 96$
\tabularnewline \hline \hline 
$\tilde{c}_3$  & -1.6211  & -1.632  & -1.649\tabularnewline \hline 
$c_4$  & 0.0196  & 0.0189  & 0.0181\tabularnewline \hline 
\end{tabular}
\caption{Perturbative values for $\tilde{c}_3$ and $c_4$}
\label{pert}
\end{table}

\noindent
We expect the values of $\tilde{c}_3$ and $c_4$, needed to restore the
symmetries of the action, to be different on the lattices we use from
the perturbative values. In this work, the tuning of these counterterm
coefficients is achieved by minimizing the parity breaking and time
asymmetry.

\begin{figure}[h]
\centering
\includegraphics[width=9cm,clip]{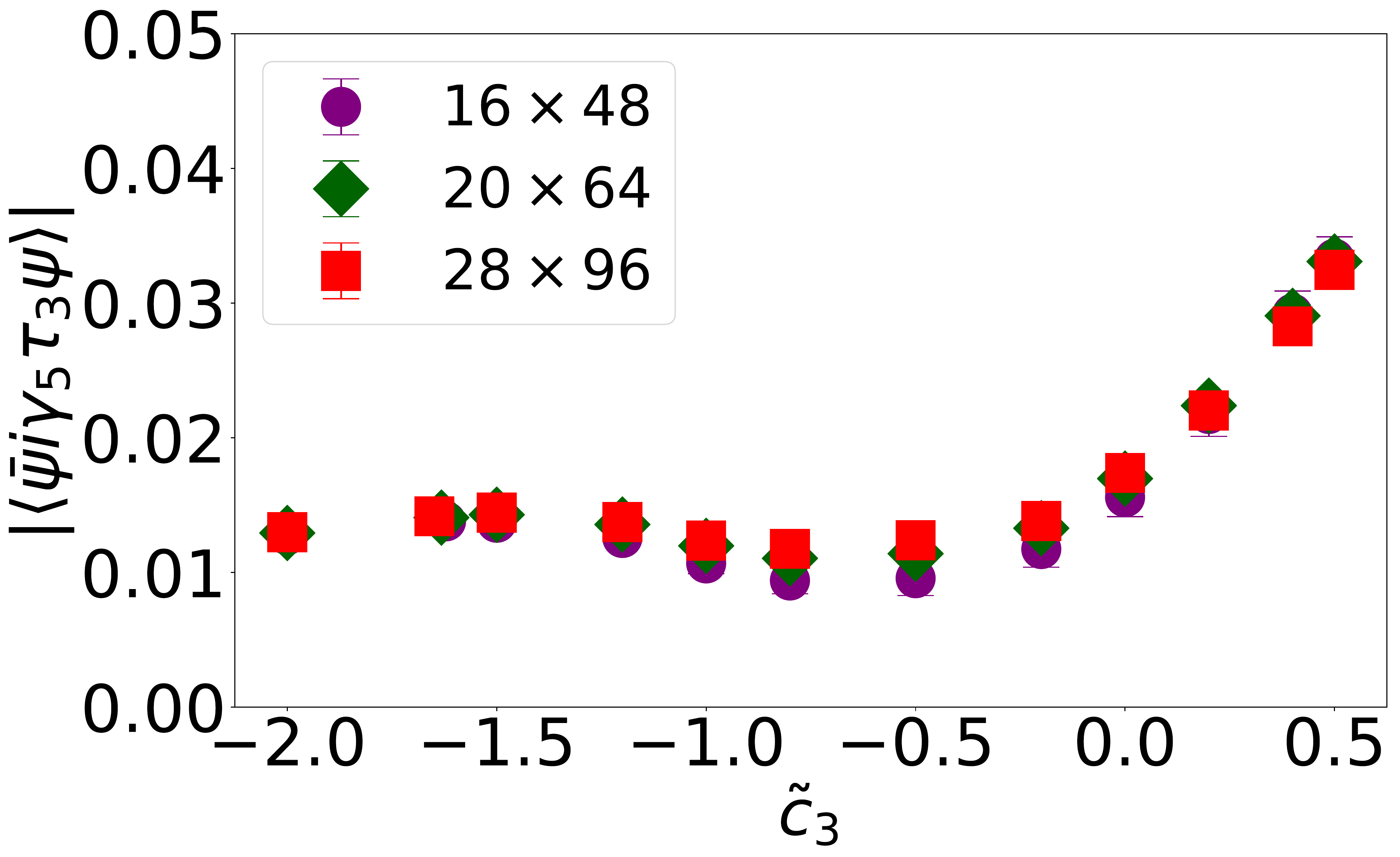}
\caption{Parity condensate versus $\tilde{c}_3$ for $c_4 = 0.0196,\,
0.0189$ and 0.0181 on $16^3\times 48,\, 20^3 \times
64$ {and} $28^3\times 96$ lattices respectively. \label{pc_c3}}
\end{figure}

The quantity we measure to determine the size of parity breaking is the
parity condensate $\vert \langle \bar{\psi} i\gamma_5\tau_3\psi\rangle
\vert$, where $\tau_3$ is the third generator of $SU(2)$. First we study
the variation of the chiral condensate as we tune $\tilde{c}_3$ keeping
$c_4$ fixed at its perturbative value for the lattice in use. The dependence
of parity condensate on $\tilde{c}_3$ for various lattices is shown in
Fig.\ref{pc_c3}. The minimum of the condensate for all three ensembles
appears around $\tilde{c}_3=-0.5$. Besides, the actual values and the
minimum of parity condensate appear to have very less dependence on the
lattice spacing.
\noindent
\begin{figure}[h]
\centering
\includegraphics[width=9cm,clip]{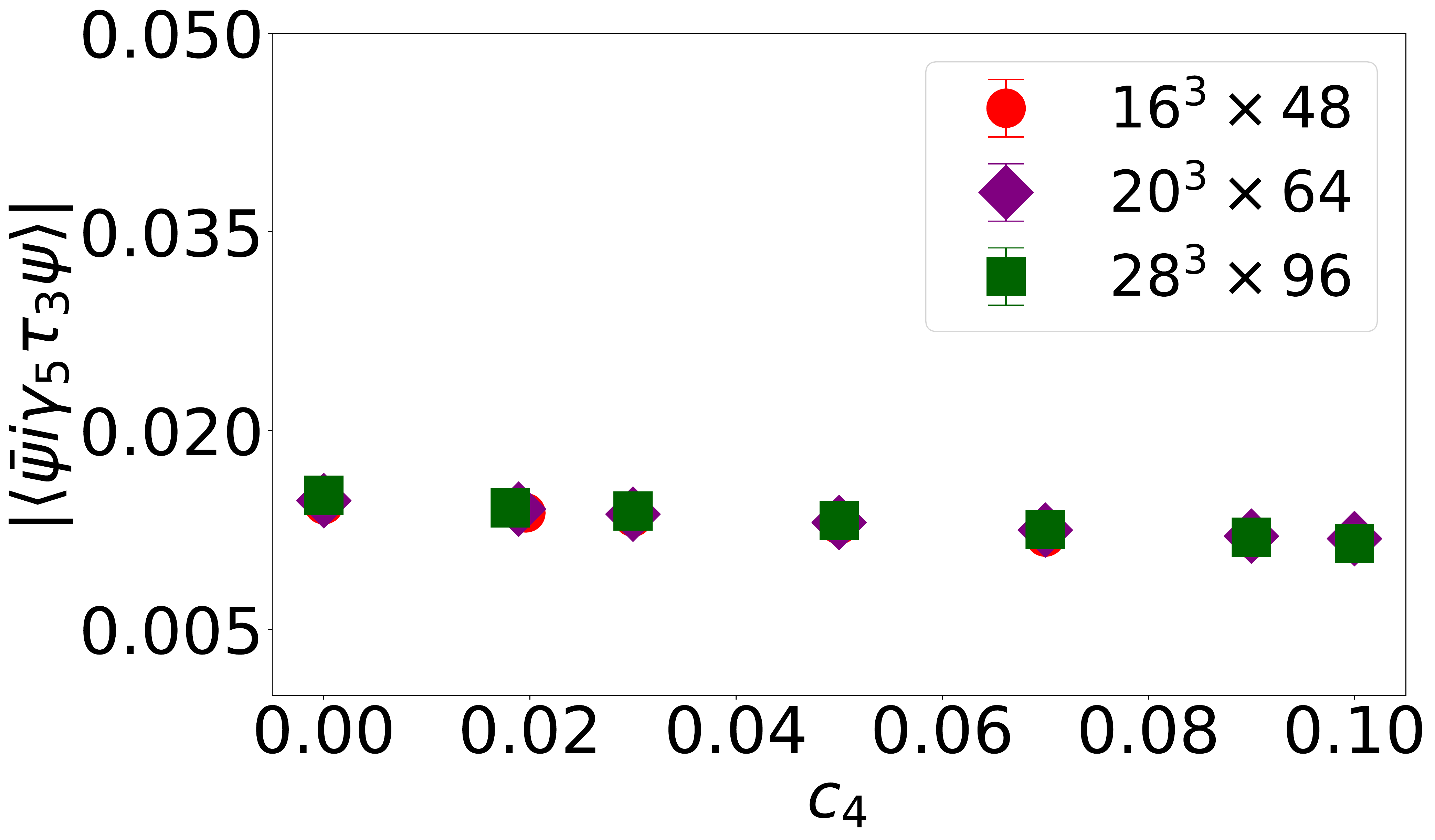}
\caption{Parity condensate versus $c_4$ for $\tilde{c}_3 = -0.5$ on 
$16^3\times 48,\, 20^3 \times 64$ and $28^3\times 96$
lattices.  \label{pc_c4}}
\end{figure}

\noindent
In the next step, we fix $\tilde{c}_3$ to this non-perturbative value
and measure parity condensate for varying $c_4$. This is plotted in
Fig.\ref{pc_c4}. We observe very little dependence of parity condensate
on $c_4$ for all the three lattices. We have checked that this nature
is also true for other values of $\tilde{c}_3$. Here too we find that
the parity condensate is almost independent of the lattice spacing.
Therefore, tuning $\tilde{c}_3$ alone is sufficient for minimizing the
breaking 
of parity.

\subsection{Time asymmetry}
The Borici-Creutz action has $PT$-symmetry and, because of breaking of
parity, we expect to see sign of time asymmetry in the theory and lattice
calculation. In lattice simulation, this time asymmetry can show up in
the spectrum and manifests through non-degeneracy of forward and backward
propagating meson states.
\begin{figure}[h]
\begin{minipage}[c]{0.98\textwidth}
\small{(a)}\includegraphics[width=7cm, clip]{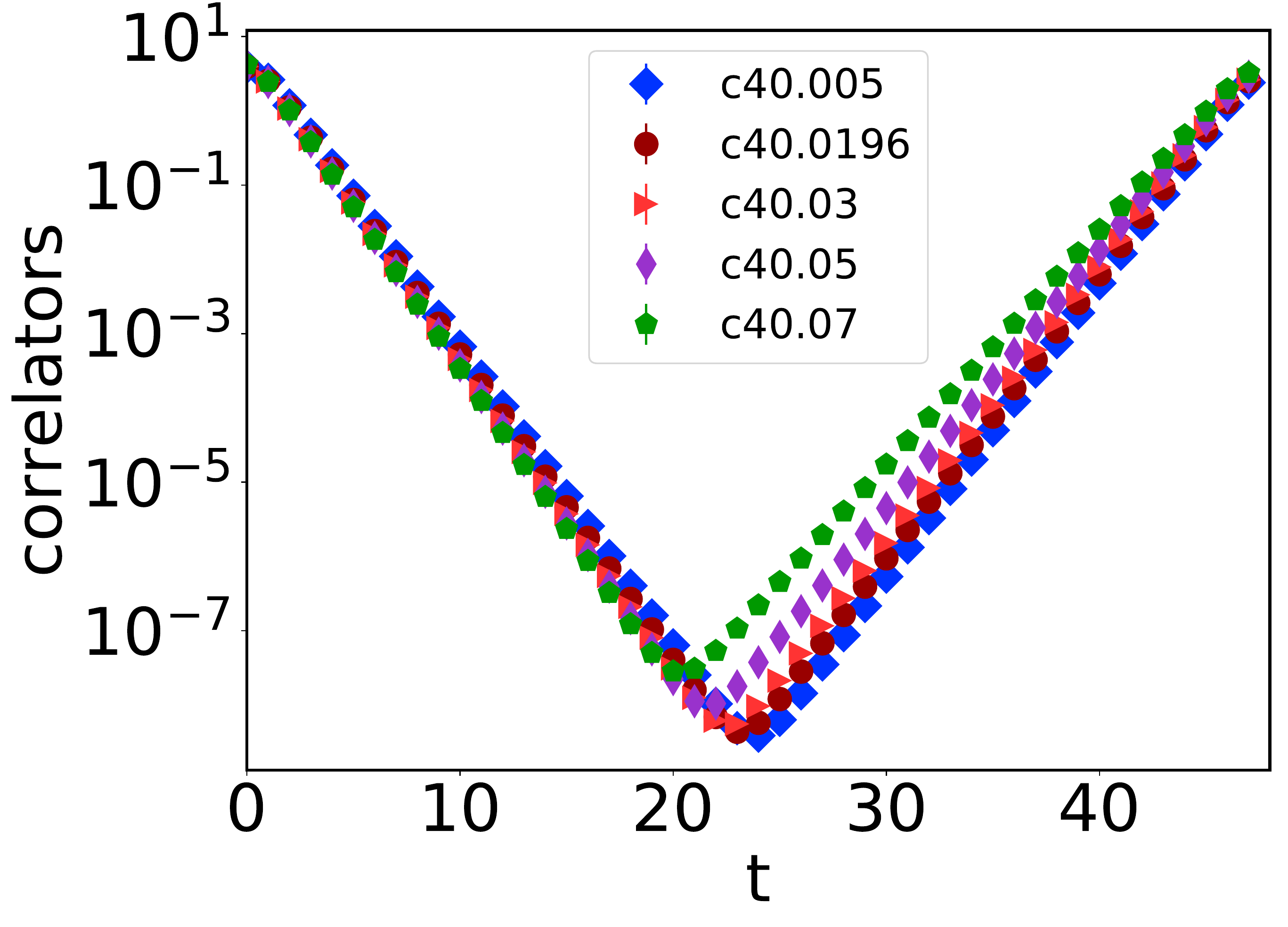}
\hspace{0.1cm}%
\small{(b)}\includegraphics[width=7cm, clip]{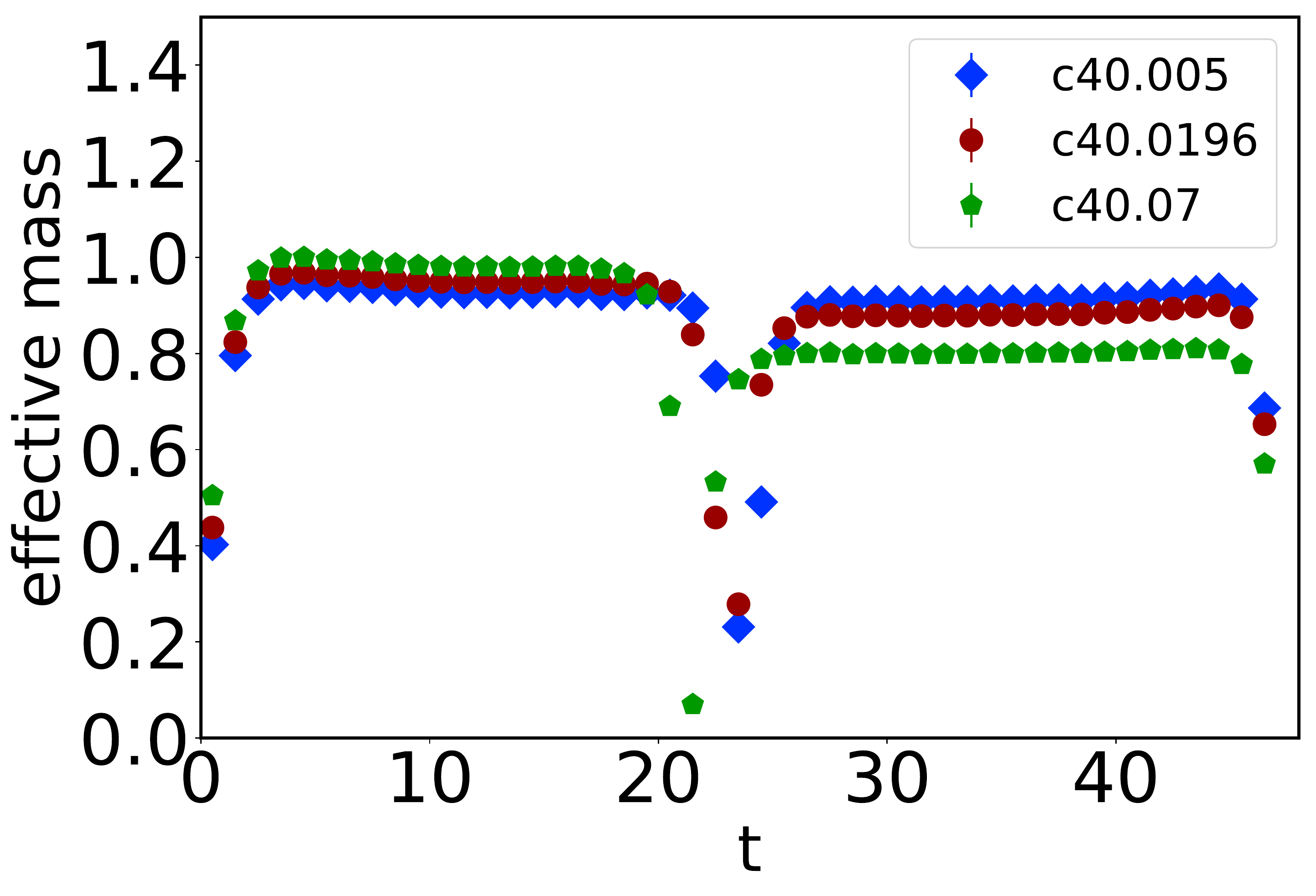}
\end{minipage}
\begin{minipage}[c]{0.98\textwidth}
\small{(c)}\includegraphics[width=7cm, clip]{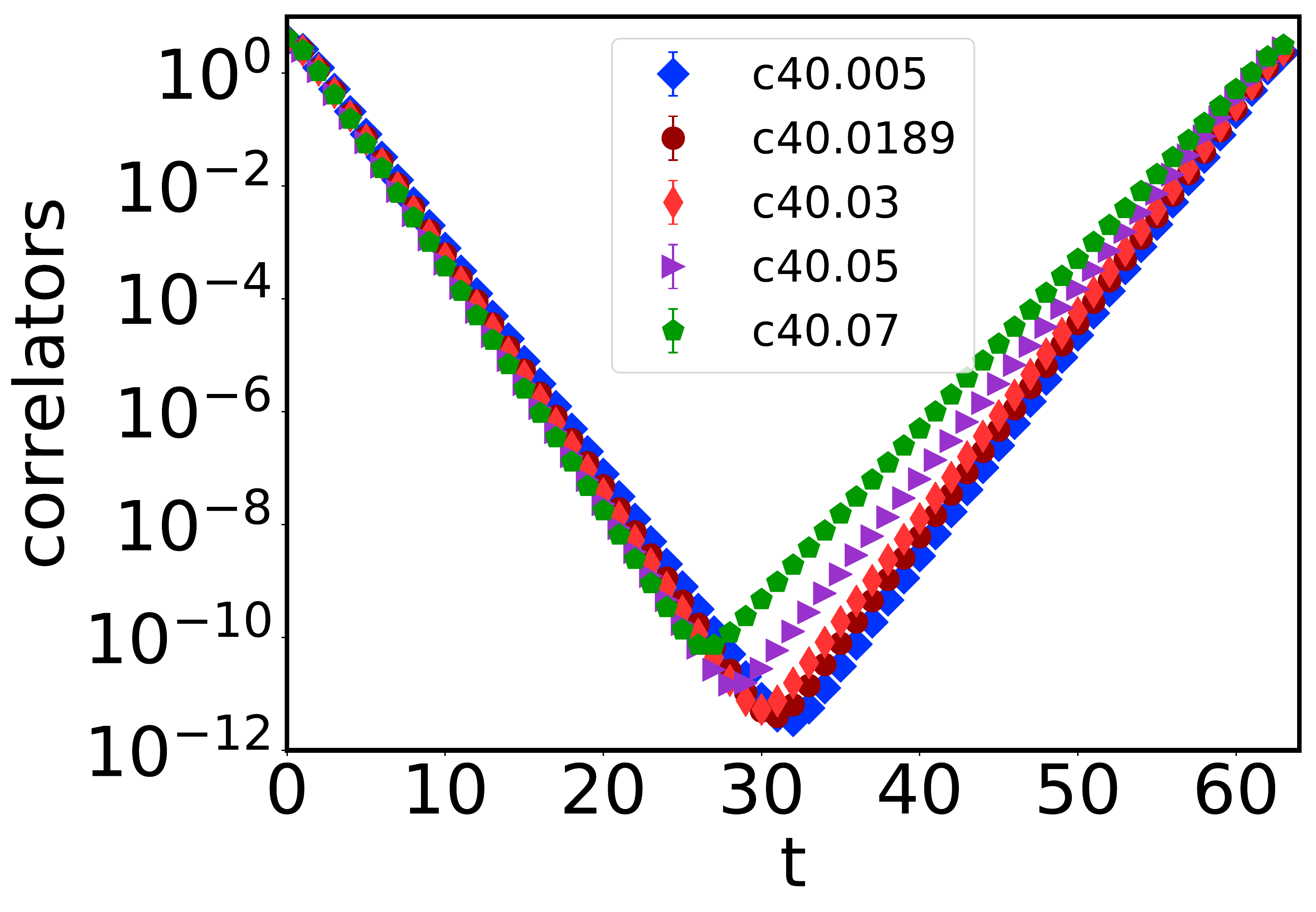}
\hspace{0.1cm}%
\small{(d)}\includegraphics[width=7cm, clip]{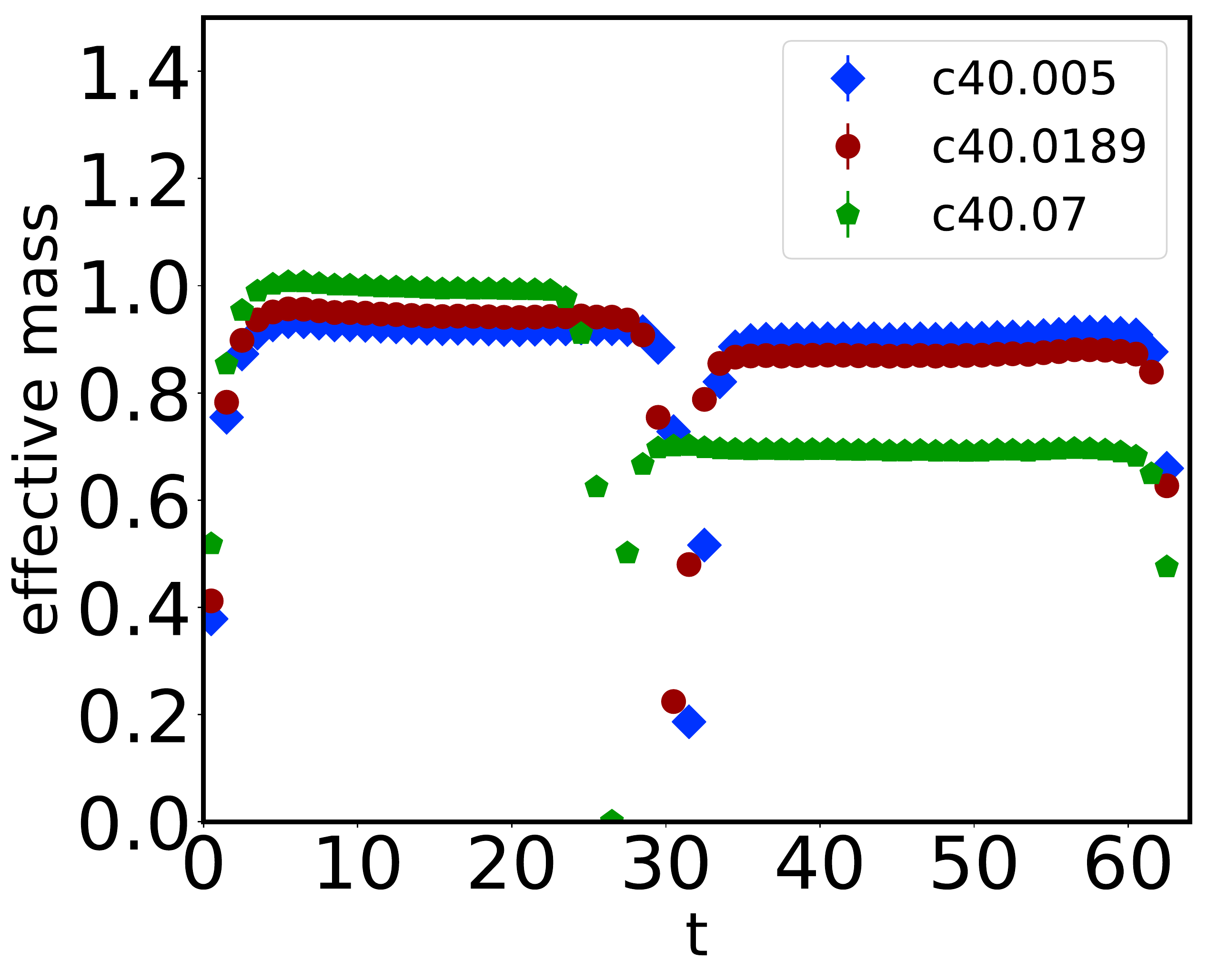}
\end{minipage}
\caption{Pion correlator and effective mass for different values of $c_4$
keeping $\tilde{c}_3 = -0.5$ fixed. Plots in upper panel (a) and (b) are
for $16^3\times 48$ lattices and lower panel (c) and  (d) are for $20^3
\times 64$ lattices. Similar results for $28^3 \times 96$ lattices are
not shown here.}
\label{T_asym}
\end{figure}

\begin{figure}[h]
\begin{minipage}[c]{0.98\textwidth}
\small{(a)}\includegraphics[width=7cm, clip]{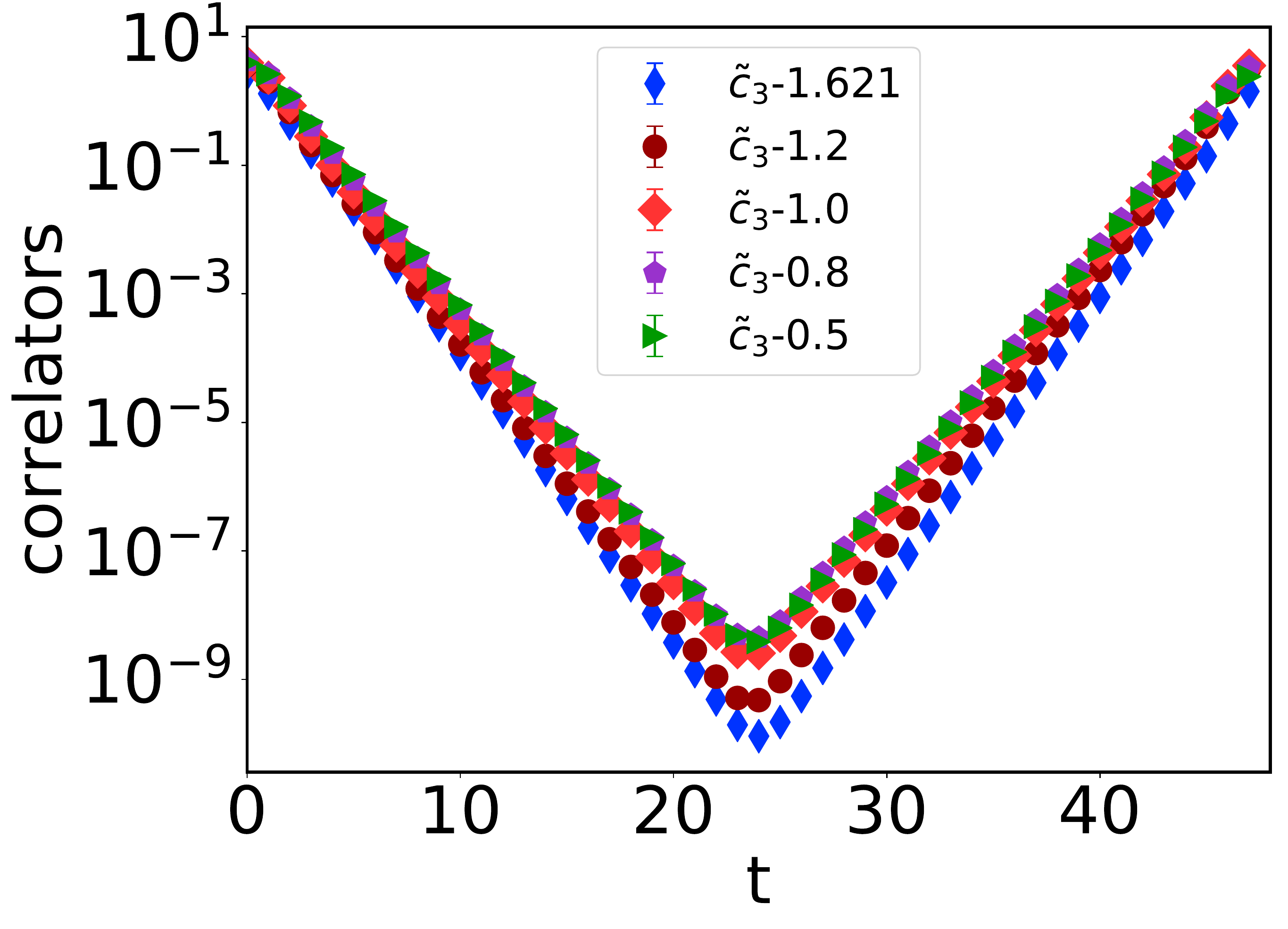}
\hspace{0.1cm}%
\small{(b)}\includegraphics[width=7cm, clip]{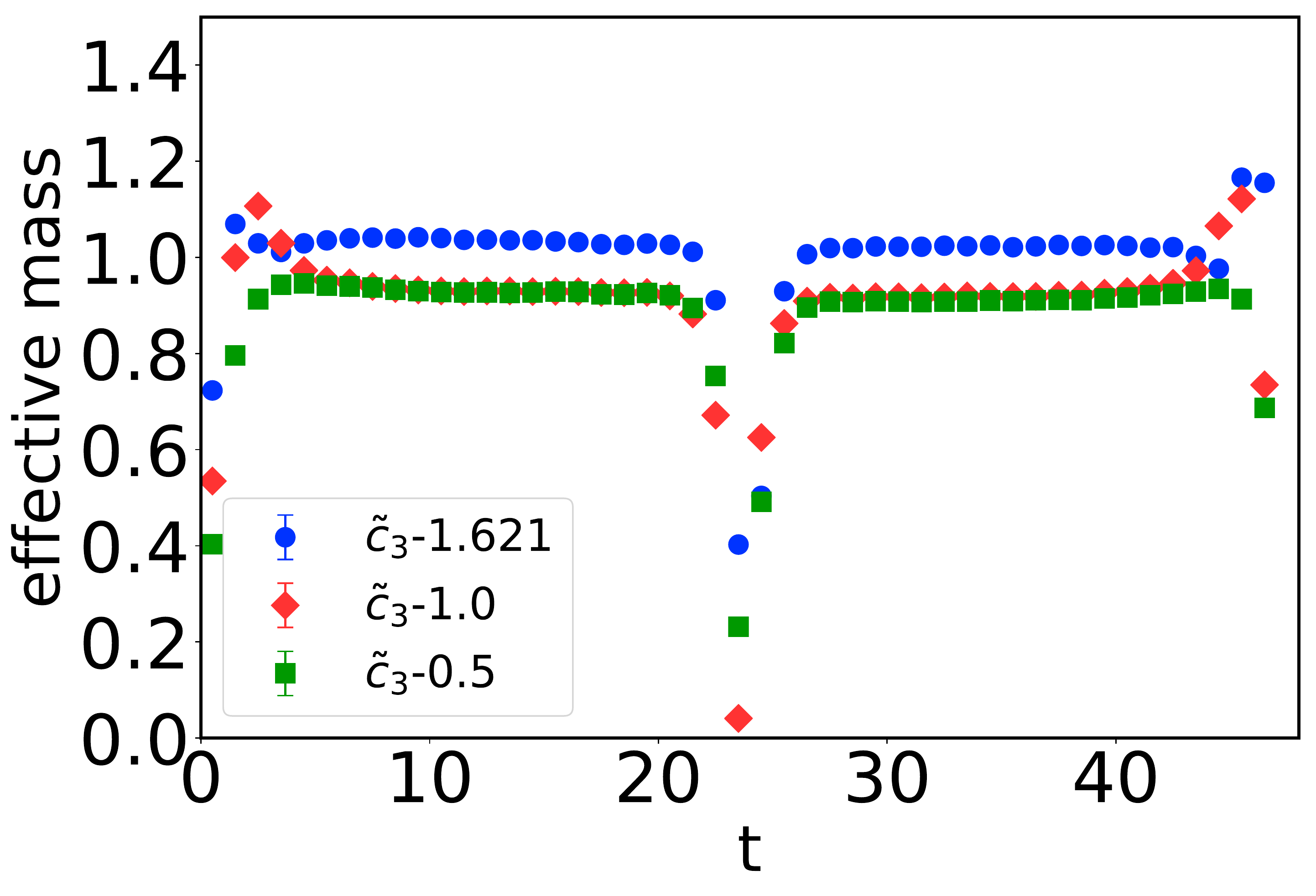}
\end{minipage}
\begin{minipage}[c]{0.98\textwidth}
\small{(c)}\includegraphics[width=7cm, clip]{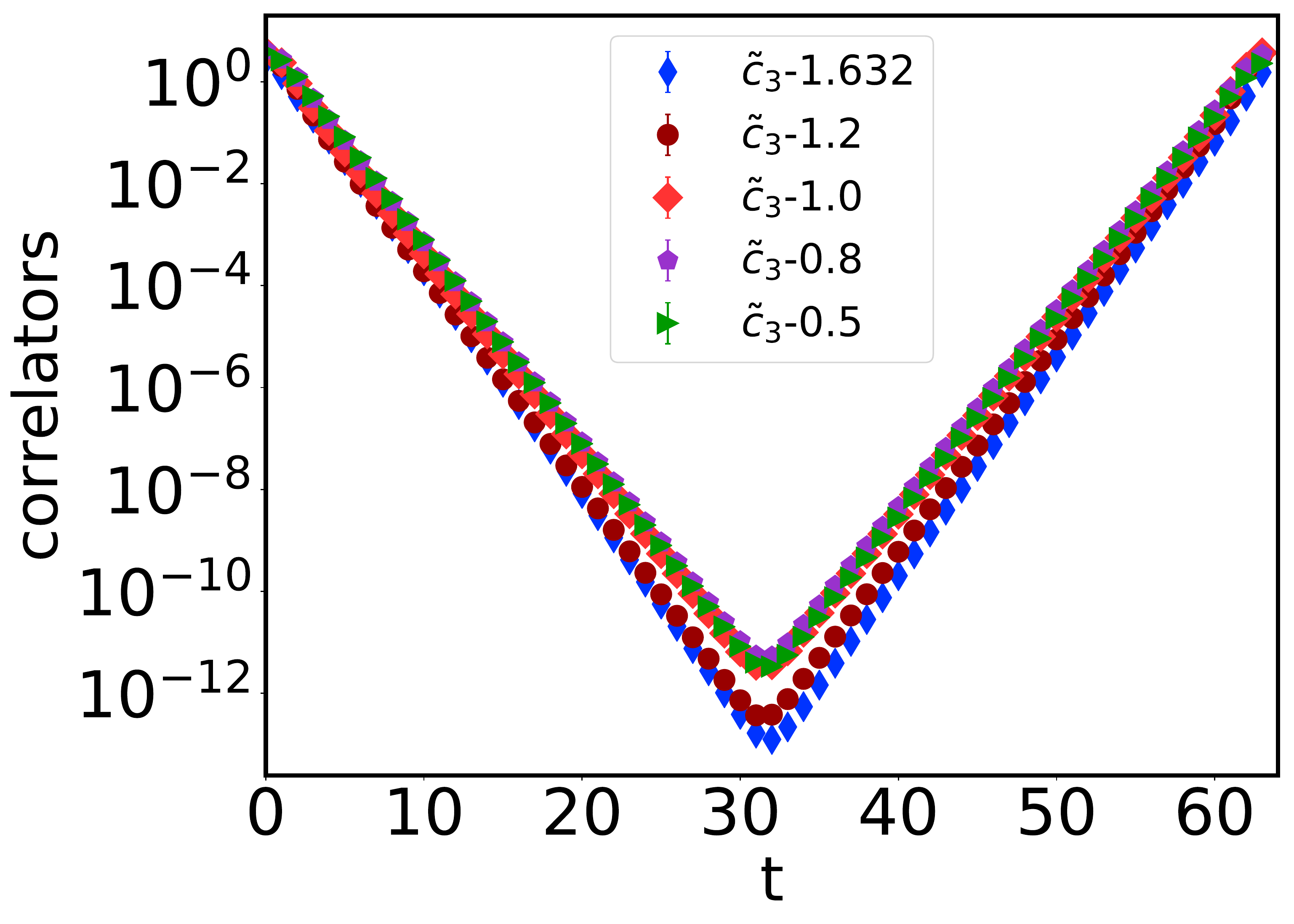}
\hspace{0.1cm}%
\small{(d)}\includegraphics[width=7cm, clip]{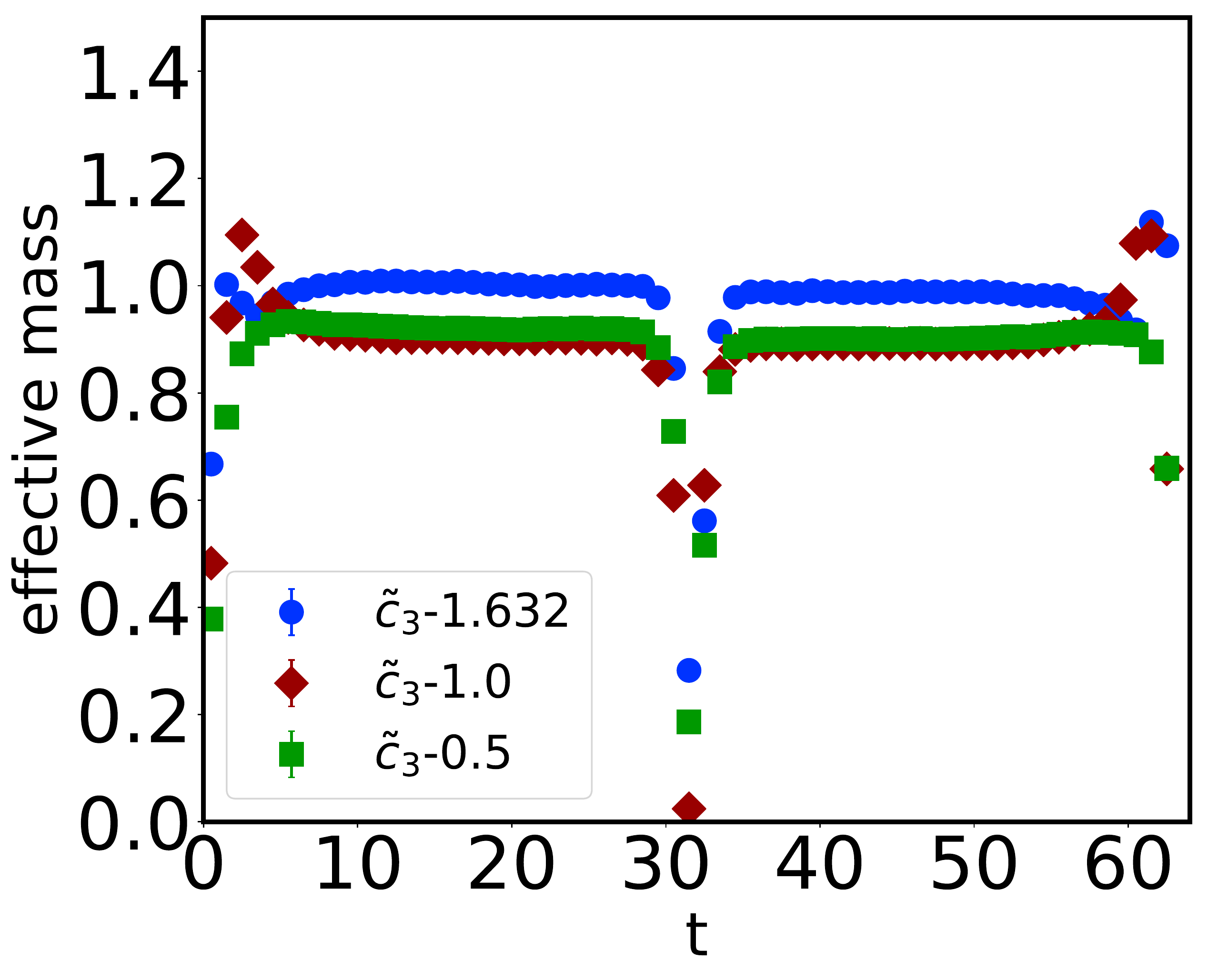}
\end{minipage}
\caption{Pion correlator and effective mass for different values of
$\tilde{c}_3$ at fixed $c_4 = 0.005$. (a) and (b) for $16^3~\times~48$
lattice, (c) and (d) for $20^3\times 64$ lattice.}
\label{T_asym_c3}
\end{figure}

The pion propagators and the corresponding effective masses for different
$c_4$, while keeping $\tilde{c}_3$ fixed at -0.5 as obtained above, are
shown in Fig.\ref{T_asym}. The pion operator is constructed from the
$d$-quark field as defined in \autoref{ud_sum}. The asymmetry in the forward
and backward propagating parts and consequently difference in effective
masses is rather evident. The asymmetry in pion propagators and the mass
difference vanishes when $c_4$ is lowered to about 0.005. This behavior
does not change with the changing values of $\tilde{c}_3$ implying that
the size of time symmetry breaking is almost entirely driven by $c_4$.
Additionally, the value of $c_4$ at which the time symmetry is restored
remain fairly constant for all the three lattices that we used.
To ascertain that $\tilde{c}_3$ has practically no role in restoring
time symmetry, we carried out the above experiment for various values of
$\tilde{c}_3$ keeping $c_4=0.005$ fixed. The plots in Fig. \ref{T_asym_c3}
show no asymmetry in pion correlators over the time slices neither
any mass difference between forward and backward propagating pions.

\begin{figure}[t]
\begin{minipage}{.99\textwidth}
(a)\includegraphics[width=7.3cm,clip]{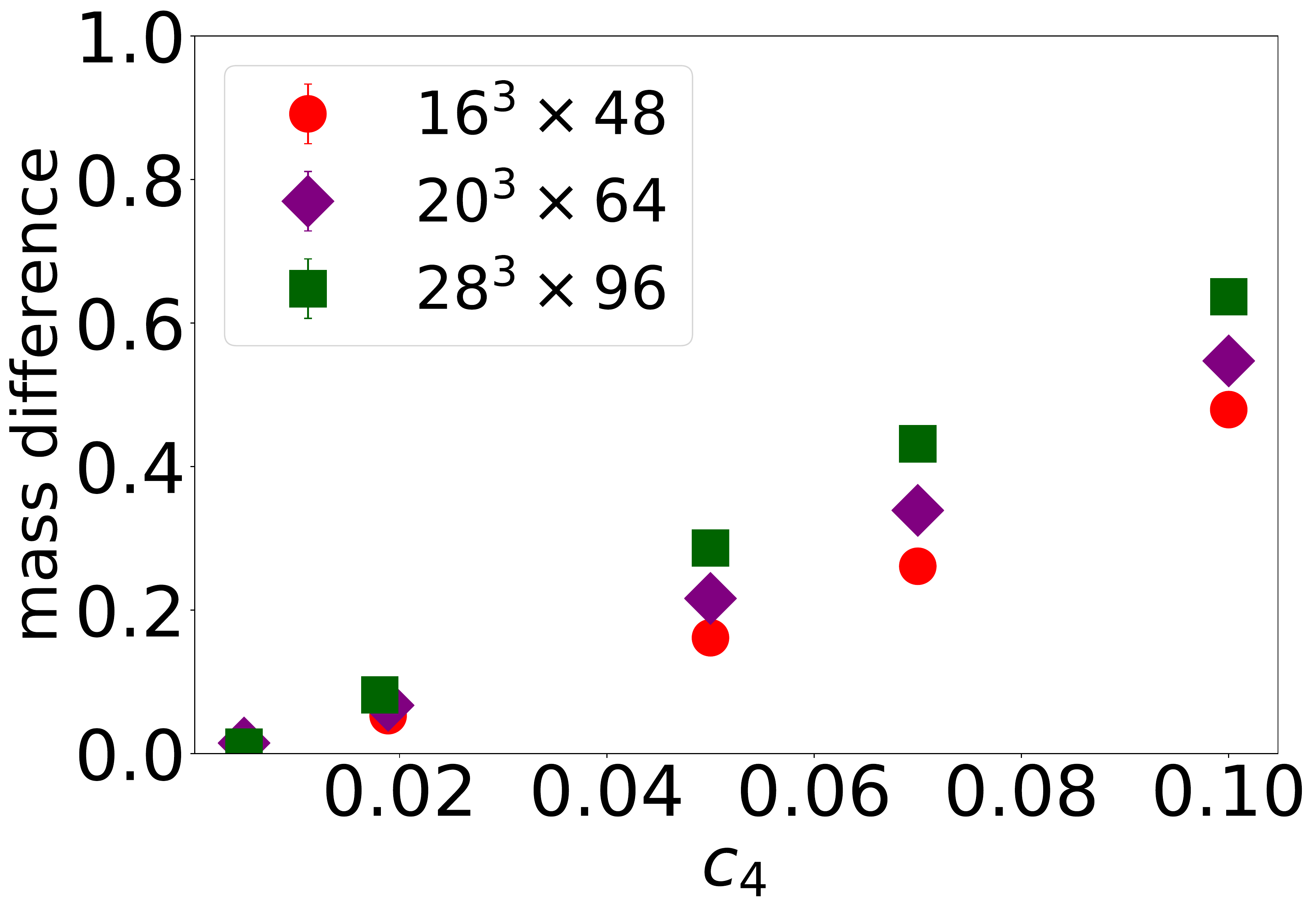}
(b)\includegraphics[width=7.6cm,clip]{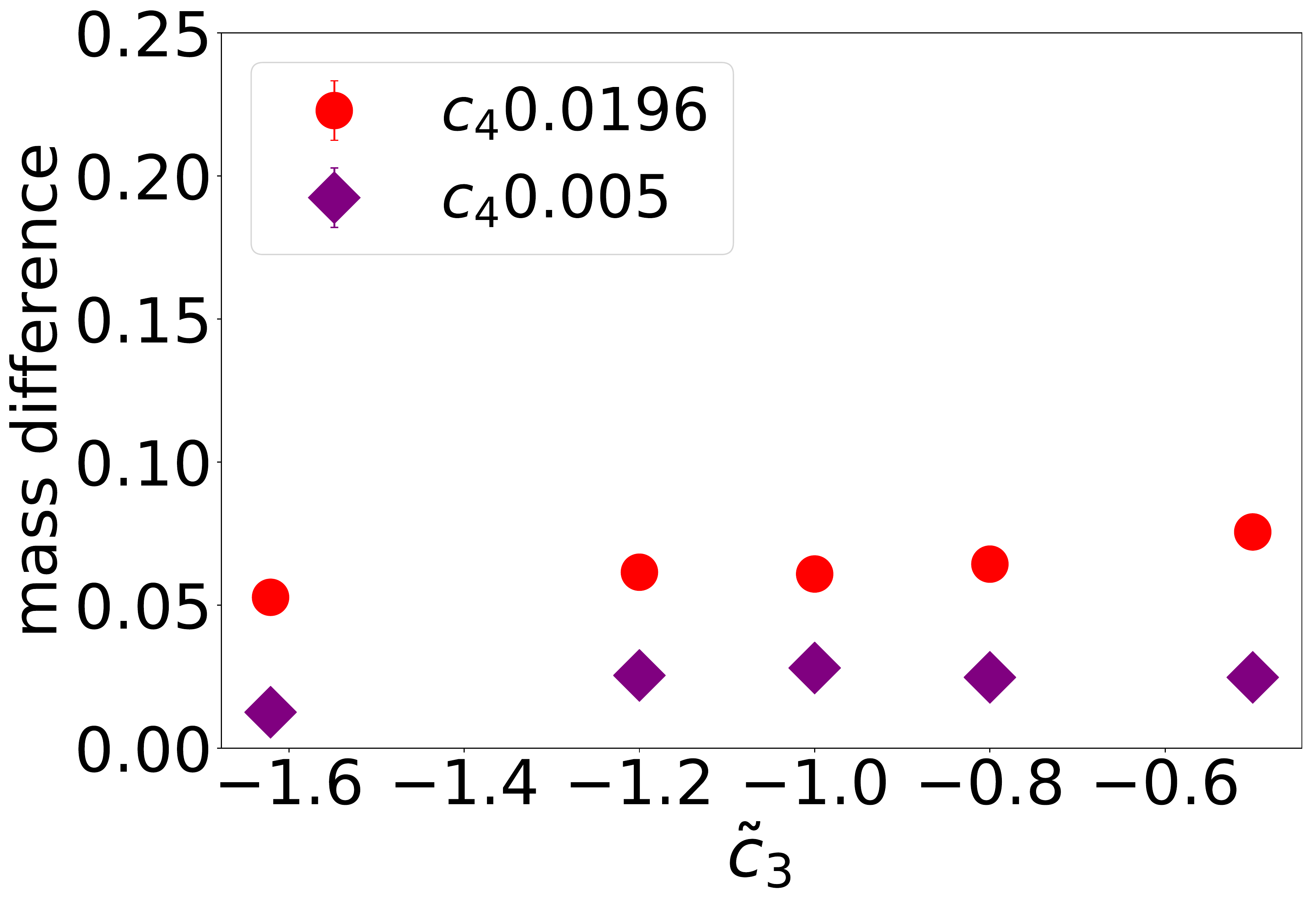}
\end{minipage}
\caption{Variations of forward-backward mass difference with the
coefficients of the counter terms. (a) Variation with $c_4$ when
$\tilde{c}_3 =-0.5$ for $16^3\times 48, \,20^3\times 64,$ and $ 28^3\times
96$ lattices. (b) Variation with $\tilde{c}_3$ for $c_4 = 0.0196,\,
0.005$ on $16^3\times 48$ lattice.
\label{massdiff}}
\end{figure}

In Fig.\ref{massdiff}, we plot the mass difference of the forward and
backward propagating pion masses against the variation of $c_4$ while
keeping $\tilde{c}_3=-0.5$ in (a) while in (b) the variation against
$\tilde{c}_3$ is shown keeping $c_4$ fixed at 0.005. Fig. \ref{massdiff}(b)
clearly indicates absence of any role of $\tilde{c}_3$ in restoring time
symmetry. Also the plot shows, although the mass differences are different
for different lattices in absence of time symmetry, the mass differences
converge to zero at about $c_4=0.005$ upon restoration of symmetry. The
smallness of $c_4$ suggests that $T$-symmetry is only weakly broken.
Besides, the values of $\tilde{c}_3$ and $c_4$ thus tuned are significantly
different from their 1-loop values as given in \autoref{pert}. Henceforth,
we have used $\tilde{c}_3=-0.5$ and $c_4=0.005$ in all our subsequent
simulations irrespective of lattice ensembles. 
The renormalization counter terms should not be  observable 
dependent, once the counterterms are tuned nonpertubatively 
which in principle include all orders of perturbation theory to restore the  
broken symmetries using certain observables,
we expect that there will be no more symmetry breaking effects 
in other obervables too. The pathologies of chiral 
perturbation theory with $P$, $T$ symmetries e.g., chiral log in pion mass or 
negative scalar correlator due to partial quenching in mixed action etc are also
observed with the renormalized BC action as expected. We have also measured 
various pseudoscalar
and vector mesons for the GMOR and $SU(6)$ mass formula, which appear
in the following section, and their behavior are consistent with
other lattice fermions. 

\section{Mixed action pion mass} 
\begin{figure}[h]
\begin{minipage}[c]{0.98\textwidth}
\small{(a)}\includegraphics[width=7cm,page=2, clip]{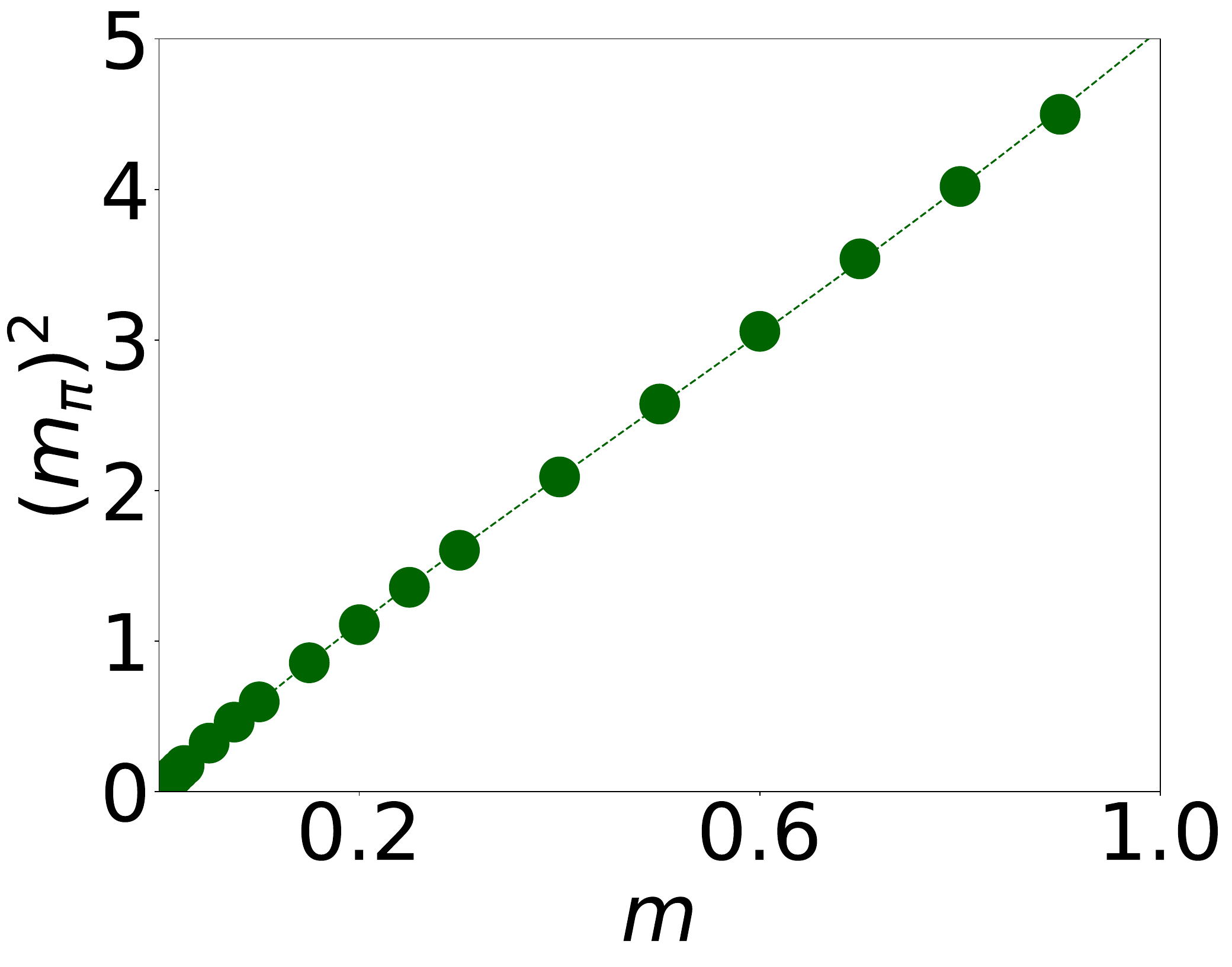}
\hspace{0.1cm}%
\small{(b)}\includegraphics[width=7cm, page=3, clip]{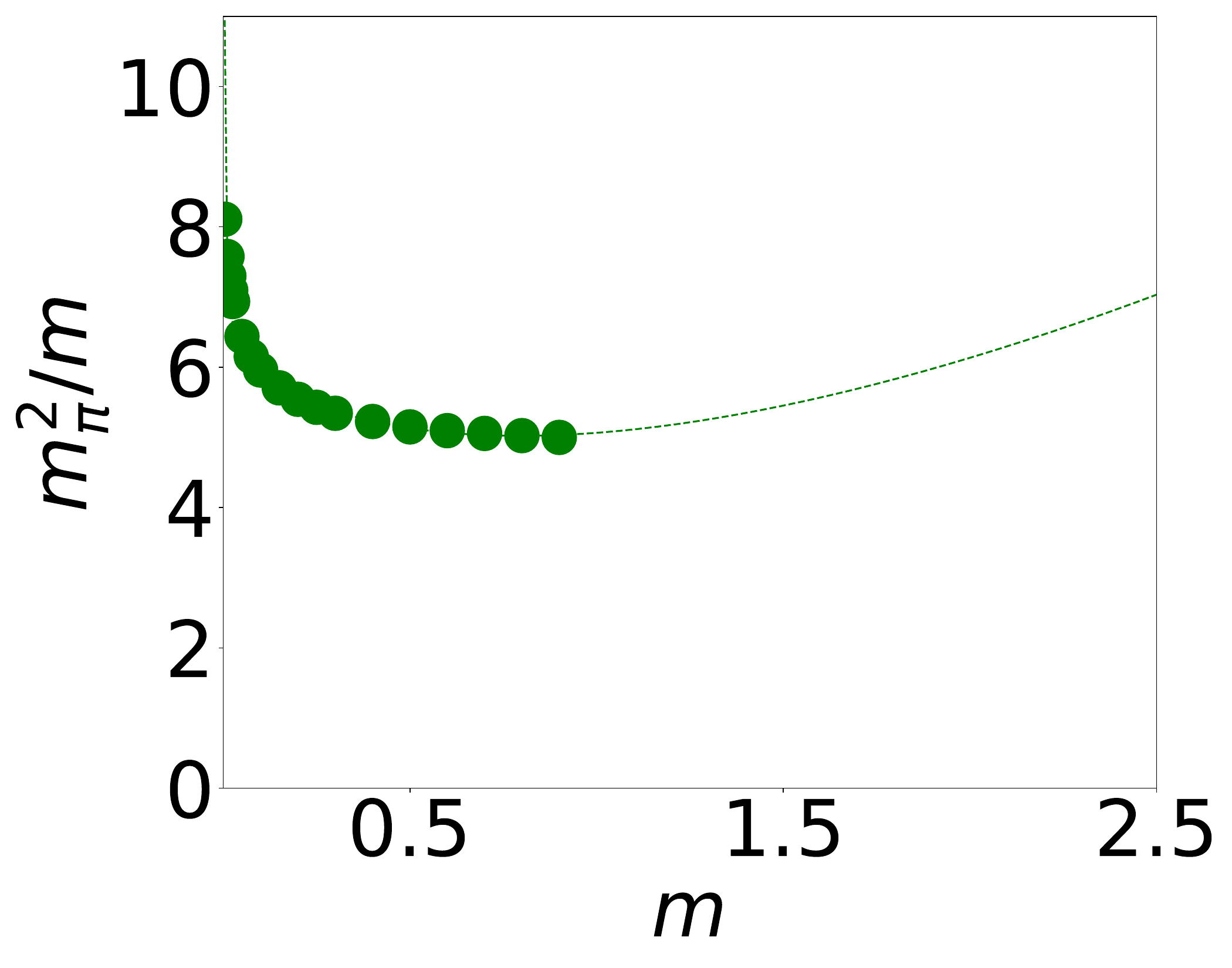}
\end{minipage}
\caption{Pion mass-squared as a function of valence quark mass $m$
on $20^3\times
64$ lattice. Error bars are smaller than the symbol size. In (b), the
dotted curve is a fit to the mass formula in \autoref{chiral_fit}.}
\label{mpi_vs_mq}
\end{figure}

Having tuned the counterterm coefficients, we next turn to the pion
spectrum. Both the $d$ and $u$ quark fields, defined in \autoref{ud_sum},
can be used to construct meson $q\bar{q}$ operators but they give
identical results. Here we exclusively quote the results for $d\bar{d}$
mesons. The pion spectrum is obtained for valence quark mass in the
range of [0.0075 -- 0.5]. In Fig.\ref{mpi_vs_mq}(a) and (b), we plotted
the pion mass $m_\pi$ and pion mass squared $m_\pi^2$ respectively
as function of bare valence quark mass $m$. At small quark masses,
the plot deviates from straight line due to the logarithmic correction
coming from one loop chiral perturbation theory,
\be 
m_\pi^2= 2 B  m + (2 B)^2 \frac{m^2}{(32 \pi^2 f^2)} \log \left(m\right)
\;\;\; , \label{1l_chipt}
\ee
where $B$ is the LEC and $f$ is the pion decay constant. The chiral log
has been studied previously for different lattice fermions in the context
of mixed action in \cite{aubin2,chen,basak}. A more complete mass formula
in one loop partially quenched  chiral perturbation theory is given by
\cite{sharpe,chen},
\be
m^2_\pi = C_1 m + C_{1L} m \log(m) + C_2 m^2 + C_{2L} m^2 \log(m),
\label{chiral_fit}
\ee
where $C_1,\,C_{1L},\,C_2,\,C_{2L}$ are independent low energy constants.
The existence of chiral log in our mixed action simulation is more
prominently visible when the ratio of $m_\pi^2/m$ is plotted
against $m$. Here we chose to show the results from only $20^3 \times
64$ lattice to avoid repetitions.

We plot in Fig.\ref{mpi_vs_mq}(b) the results for $m_\pi^2/m$
versus $m$ for the same range of quark mass to highlight the evidence
of the partially quenched chiral logarithm due to mismatch of valence
and sea quark masses. In the same plot we show the fitting of our
results with the PQ$\chi$PT formula of \autoref{chiral_fit}. Here,
however, away from the smaller quark mass the $m_\pi^2/m$
attains some sort of plateau and is expected to rise at still higher
masses.

\begin{figure}[htp]
\begin{minipage}{.98\textwidth}
\includegraphics[width=7.2cm,clip]{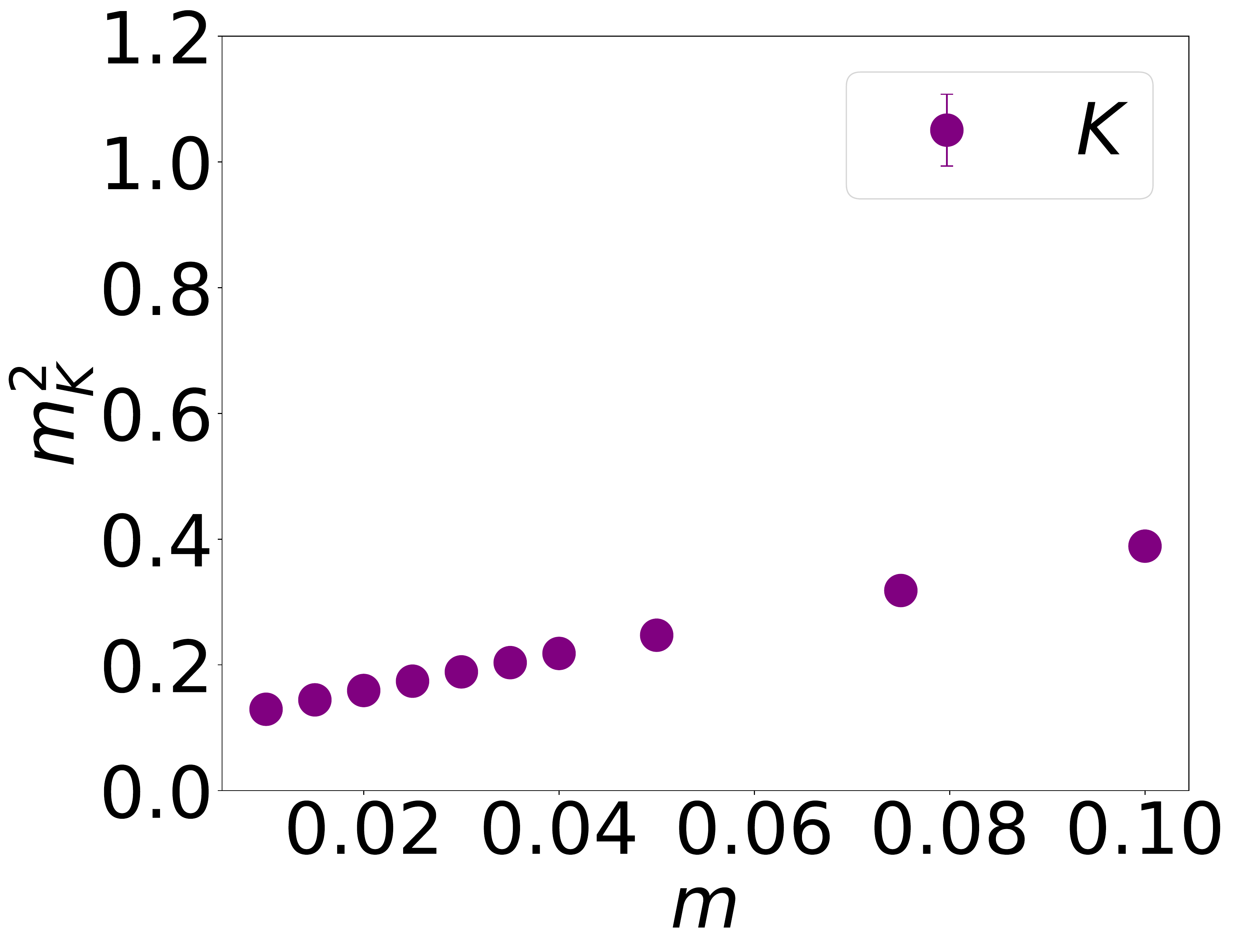}
\hspace{0.1cm}%
\includegraphics[width=7.2cm,clip]{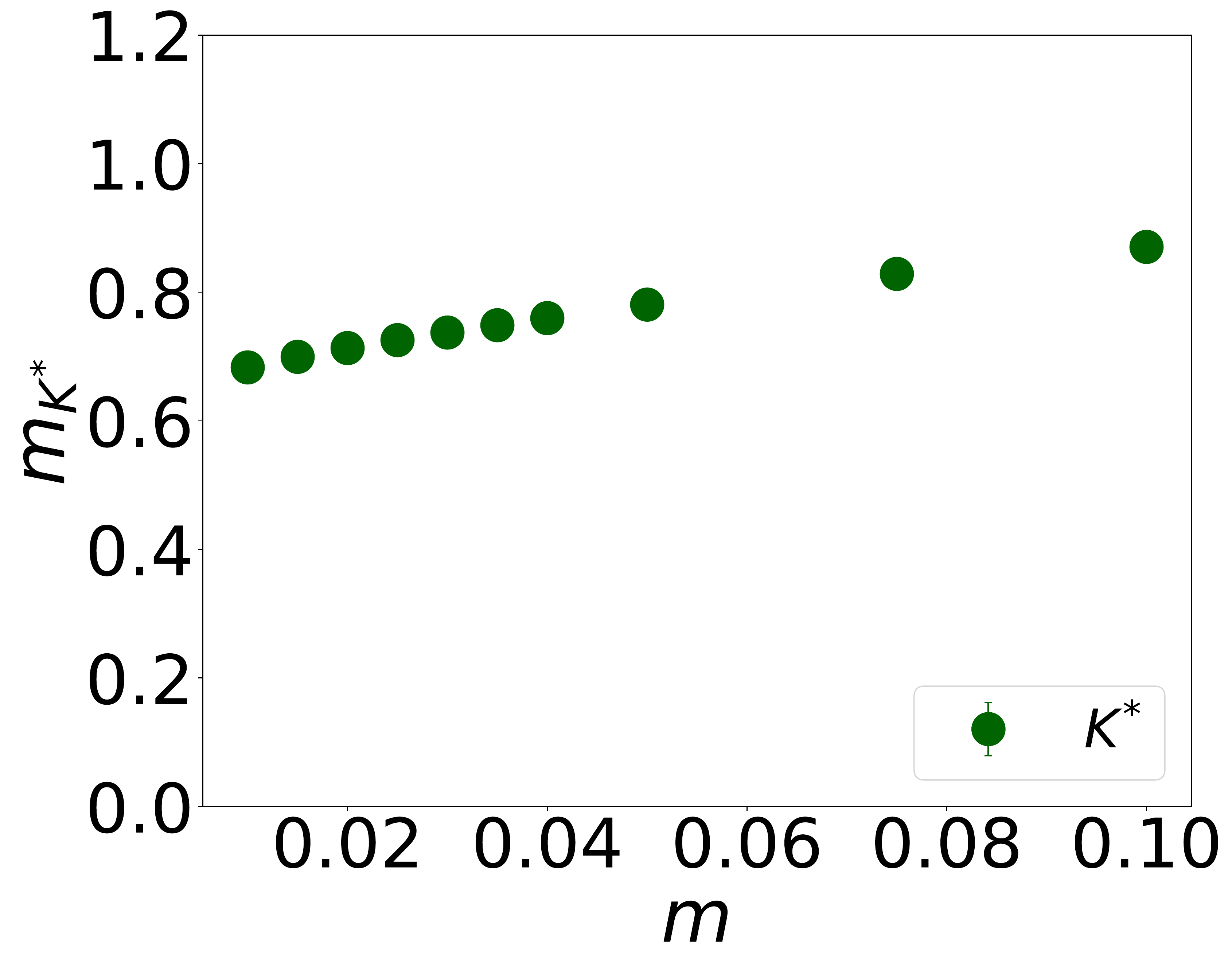}
\end{minipage}
\caption{\label{kaon_mass} $m_K^2$ and $m_{K^*}$ versus light quark mass on $20^3\times 64$ lattice.}
\end{figure}
Generally, the strange quark mass is determined by tunning the
unphysical pseudoscalar $s\bar{s}$ meson mass to $687$ MeV. Having
the same quantum number, $s\bar{s}$ mixes with  $u\bar{u}$ and
$d\bar{d}$ to produce $\eta$ and $\eta^\prime$ mesons. But purely
$s\bar{s}$ meson is obtained by omitting the quark-antiquark
annihilation from the simulation\cite{Davies,MILC}.
The strange mass $am_s=0.030$ gives the fictitious pseudoscalar $s\bar{s}$
mass $m_{s\bar{s}}=682$ MeV (with $a=0.13$ fm) which corresponds to the
bare strange mass $m_s=45.46$ MeV on $20^3\times 64$ lattice.
With that fitted strange mass, the vector meson $\phi=s\bar{s}$ is found
to have $m_\phi\approx 1120$ MeV[PDG\cite{pdg} value 1020 MeV].
According to leading order chiral perturbation theory, vector kaon mass
($m_{K^*}$)  and the square of the  pseudoscalar kaon mass ($m_{K}^2
=m_{K^\pm}^2=m_{K^0}^2$) depend linearly on the quark mass:
$m_K*=\lambda+\lambda_2(m+m_s)$ and $m_K^2=B(m+m_s)$.
In Fig.\ref{kaon_mass}, $m_K^2$ and $m_{K^*}$ for the above mentioned
strange mass are plotted against the lighter quark
masses. The lightest  pion mass in the plot is $m_\pi\approx 400$ MeV
(corresponding to the light quark mass $am=0.01$). For this pion mass,
$\rho$ mass is found to be $m_\rho=912$ MeV and the kaon masses on
this coarse lattice are obtained as $m_K\approx 550$ MeV[PDG value
496 Mev] and $m_{K^*}\approx 1035$ MeV[PDG value 892 MeV].

\begin{figure}[htp]
\includegraphics[width=9cm,clip]{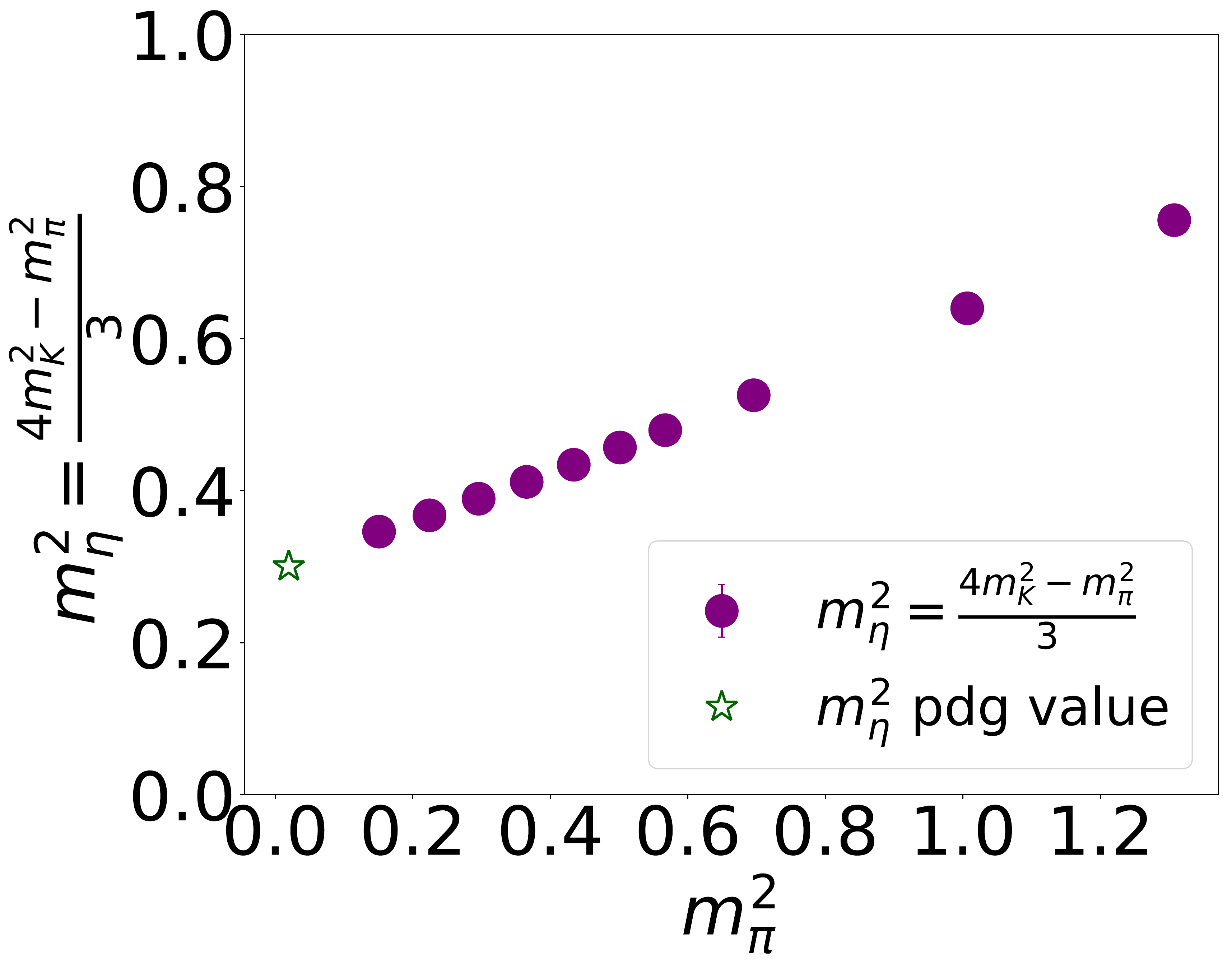}
\caption{\label{eta} The $\eta$ mass from GMOR relation on
$20^3\times 64$ lattice. The star symbol ($\star$) in the figure
indicates the PDG value of $m_\eta$ at $m_\pi=140$ MeV.}
\end{figure}
\begin{figure}[htp]
 \begin{minipage}{.98\textwidth}
 	(a)	\includegraphics[width=7cm,clip]{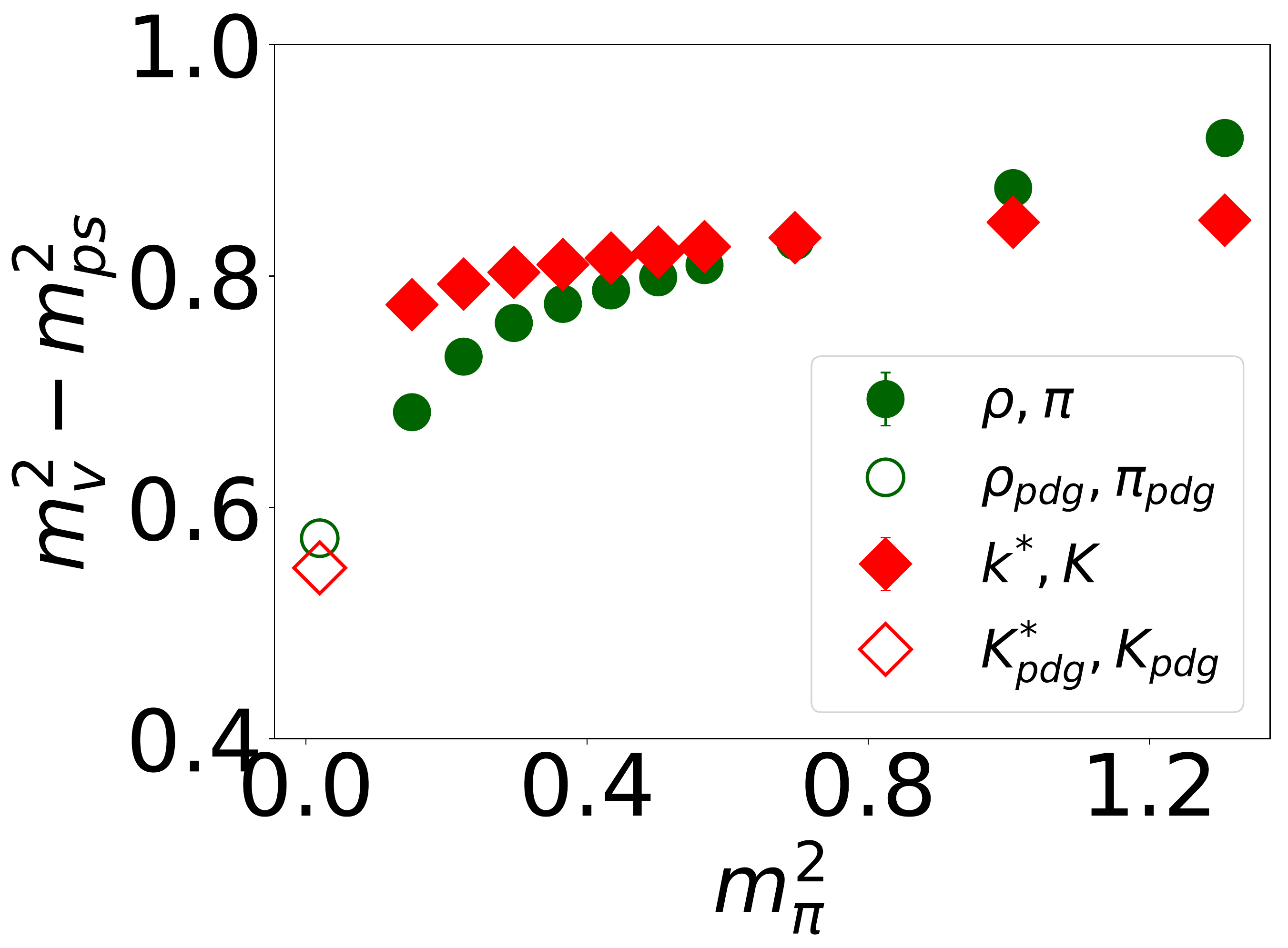}
 	\hspace{0.1cm}%
 	(b)\includegraphics[width=6.5cm,clip]{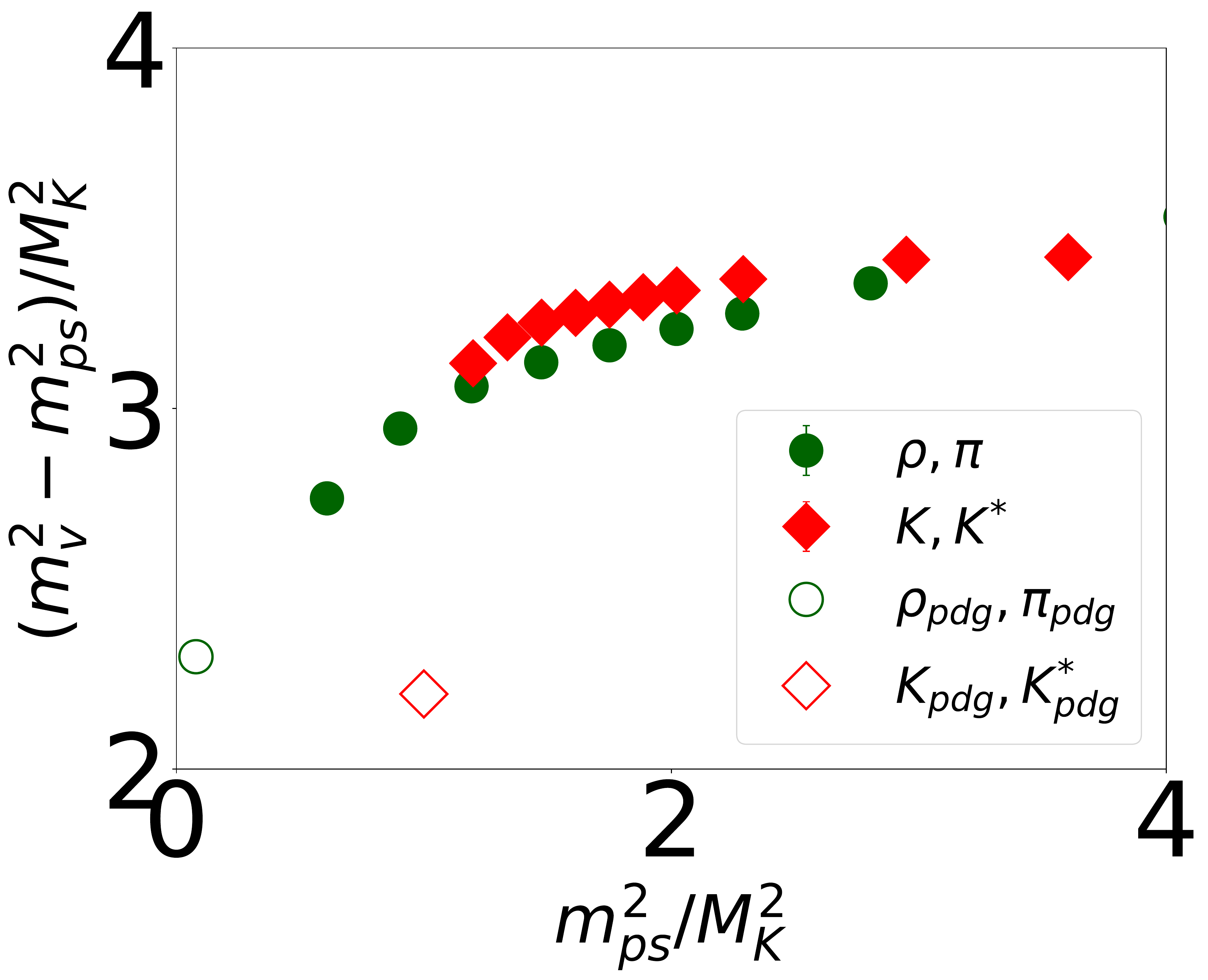}
 \end{minipage}
\caption{\label{su6_mass} (a) The vector-pseudovector mass splitting
$m_V^2-m_{PS}^2$ is plotted with $m_\pi^2$.  The $SU(6)$ mass formula is
approached with decreasing pion mass (PDG values of the splittings
are indicated by open circle and open diamond). 
(b) The ratio $(m^2_V-m_{PS}^2)/M_K^2$ is plotted
with $m_{PS}^2/M_K^2$ where $M_K$ is the PDG value of the kaon mass.
The result is from $20^3\times 64$ lattice.}
\end{figure}

A consequence of the approximate chiral symmetry in QCD is the 
Gell-Mann-Oakes-Renner(GMOR) relation. In the leading order of
chiral perturbation theory, the GMOR relation translates into
the Gell-Mann Okubo mass formula for pseudoscalar mesons which
can be written as 
\be
3m^2_\eta=4 m^2_K-m_\pi^2.
\ee 
In Fig.\ref{eta}, we have shown the GMOR relation for different pion
masses. 
Another interesting mass formula which is also found to be reasonably
satisfied by the meson spectrum  was obtained in the $SU(6)$ theory.
The $SU(6)$ mass formula \cite{su6} relates the vector-pseudoscalar
splittings as:
\be
 m_{K^*}^2-m_K^2=m_\rho^2-m_\pi^2.\label{v-ps}
\ee
The above $SU(6)$ relation was shown to deviate on lattice with
increasing quark mass\cite{su6_lat}. In Fig.{\ref{su6_mass}},
we have shown the variations of the vector-pseudoscalar splittings
with the pion mass.  We have also plotted the
ratio $(m_V^2- m_{PS}^2)/M_K^2$ for strange and non-strange sectors
 against  $m_{PS}^2/M_K^2$ where $M_K$ is the
 PDG value of the kaon mass, 
$m_V$ is $m_{K^\star}$ and $m_\rho$ when $m_{PS}$ is $m_K$ and $m_\pi$
respectively.

\section{Partial quenching and the scalar correlator}
\begin{figure}[h]
\begin{minipage}{0.98\textwidth}
 \includegraphics[width=7cm,clip]{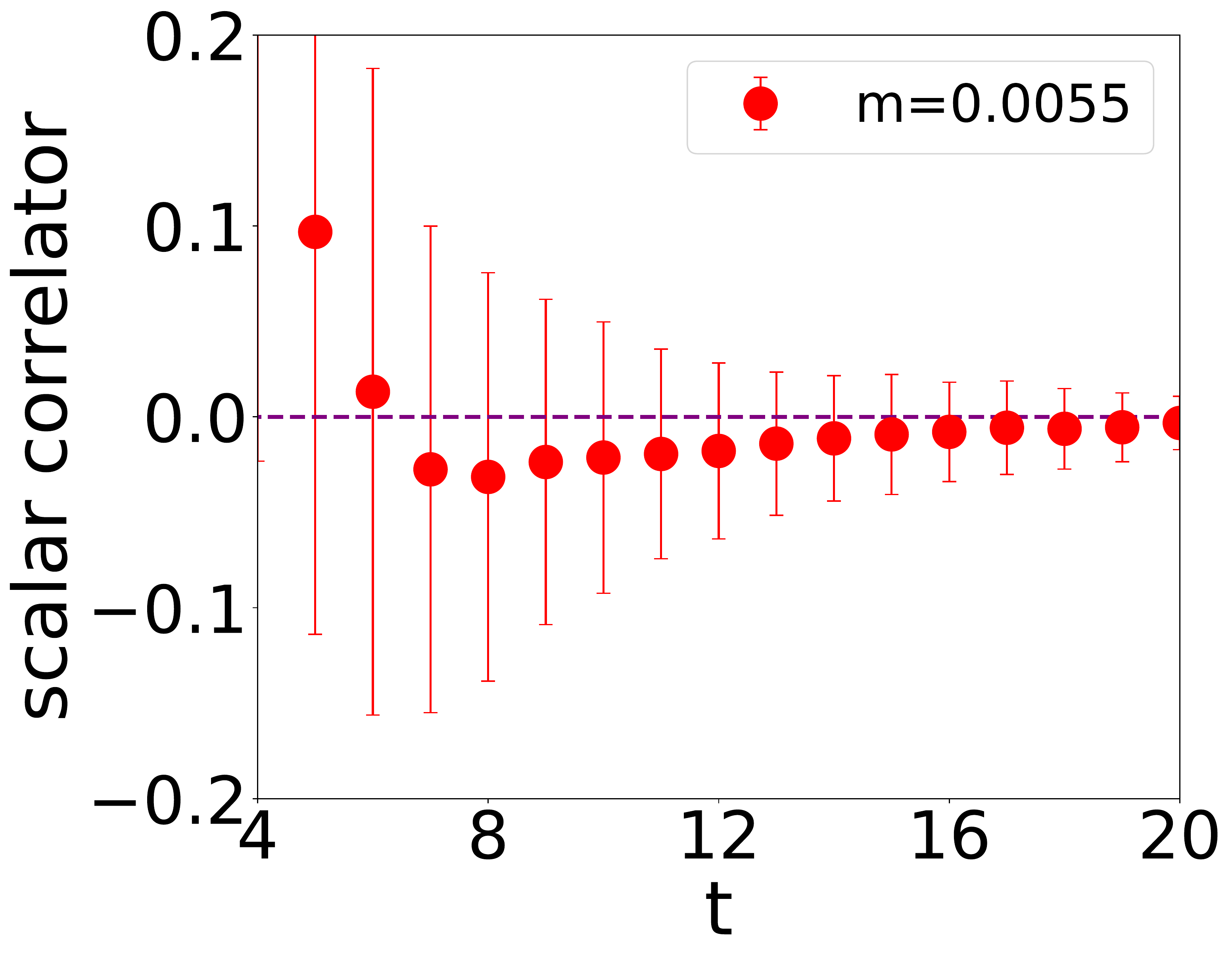}
 \hskip.2cm
  \includegraphics[width=7cm,clip]{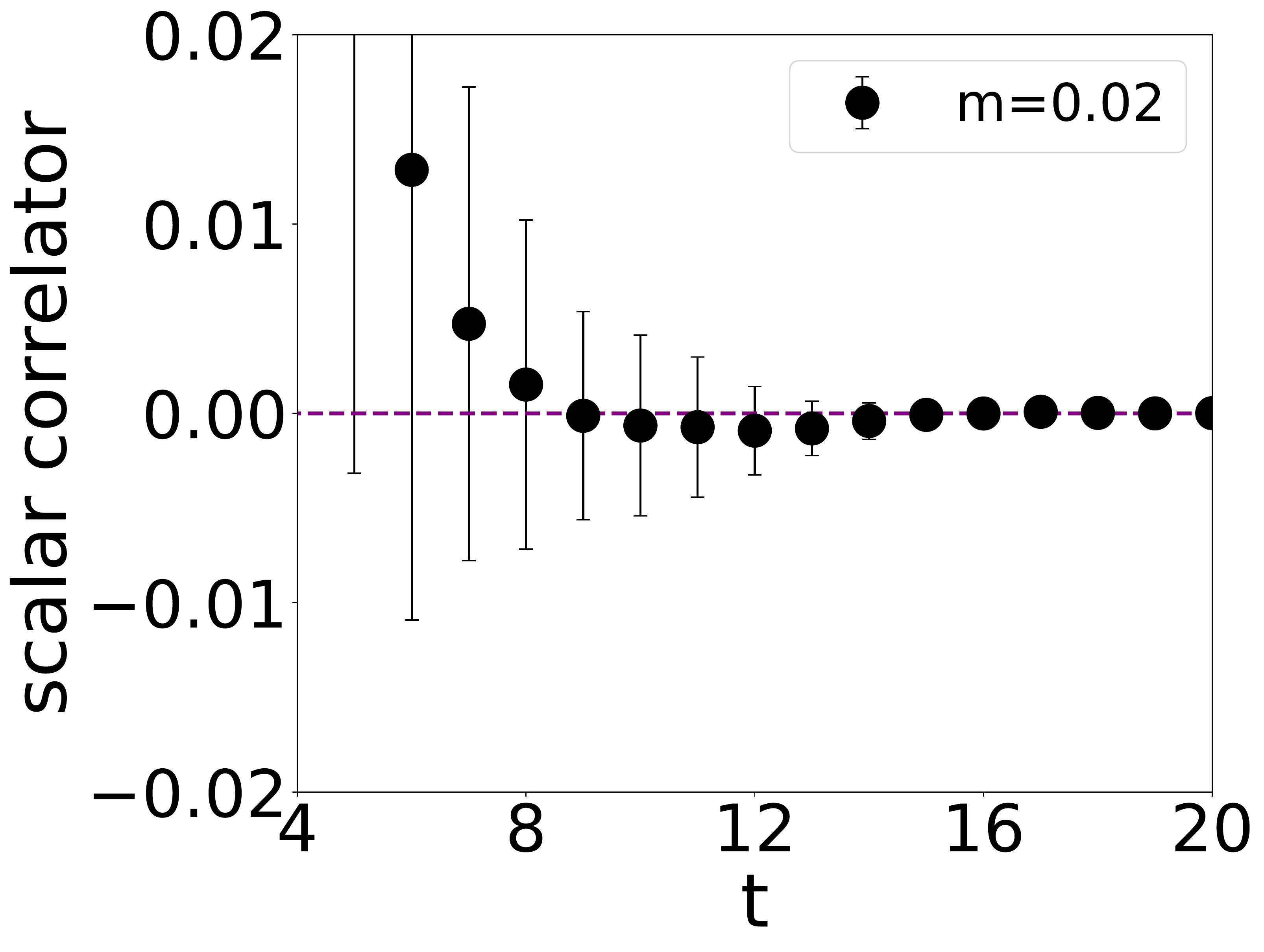}
\end{minipage}
\caption{ Scalar correlator for two different valence masses  $m_{\rm 
val}=0.0055$ and $m_{\rm val}=0.02$ 
with sea mass $m_{\rm sea}=0.01$, for nonperturbatively tuned value of the counter term $\tilde{c}_3=-0.05.$}
\label{scalar_corr}
\end{figure}
Lattice QCD with mixed action is inevitably partially quenched, no
choice of valence quark mass can completely remove unitarity violating
effects from mixed action theory at nonzero lattice spacing. The full
QCD can be recovered in the continuum limit only. The scalar meson
($\bar{\psi} \,\psi$) is known to be sensitive to this unitary
violation, it gives the scalar correlator a negative value when
valence quark mass is less than sea quark. The two point correlator
should be positive for dynamical fermions in full QCD where unitarity
is preserved. But in the partial quenched QCD when $m_{\rm val} <
m_{\rm sea}$, the flavor neutral two-meson intermediate state couples
to the scalar meson correlator with a negative weight. However, when
$m_{\rm val} > m_{\rm sea}$, the one loop contribution to the scalar
correlator coming from the exchange of the two-meson fields becomes
positive \cite{soni}.
\begin{figure}[h]
\includegraphics[width=8cm,clip]{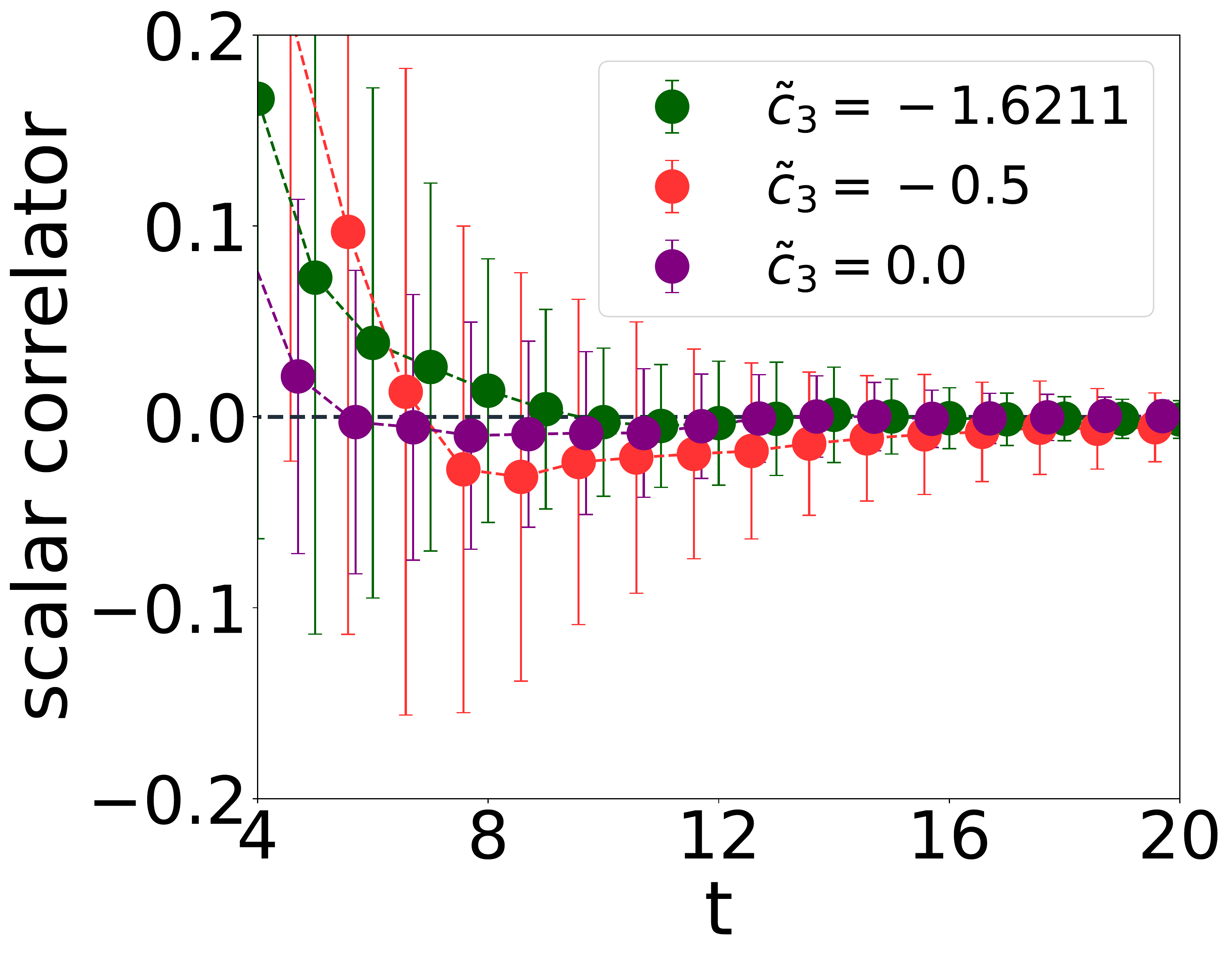}
\caption{Scalar correlator for different values of $\tilde{c}_3$ for $m_{\rm val}<m_{\rm sea}$. The lines connecting the points are to guide the eyes.}
\label{scalar_corr2}
\end{figure}

This effect of partial quenching is shown in Fig. \ref{scalar_corr},
which is observed when 
the parity breaking is minimized by non-perturbative
tuning of $\tilde{c}_3$. The errors become large for partially quenched
scalar correlators, but its negative value when $m_{\rm val} (=0.005)
< m_{\rm sea} (=0.01)$ is observable. From the plot for $\tilde{c}_3=
-0.5$, it is evident that as one increases valence quark mass, for a
fixed sea quark mass, the negative contribution reduces. By looking
at the sign of the scalar correlator as the mass of valence quark is
changed, it should be possible to match the valence and sea quark masses.
In the plots for $\tilde{c}_3$ away from non-perturbative value, {\em
i.e.} in presence of parity breaking we do not see this effect of
negative contribution.
In Fig.\ref{scalar_corr2}, the scalar correlators for different values of 
$\tilde{c}_3$ are compared to demonstrate that the partial quenching sets in when the counter term restores 
the parity.
On the question whether we can also have negative scalar 
		correlators in the parity broken phase, we cannot be definitive at this stage since relevant mixed action ChiPT involving BC fermions on staggered
		sea has not been worked out as yet. All we can emphasize at the
		moment is that our numerical data shows, in spite of large errorbars,
		in broken symmetric phase the mean of the scalar correlators are
		never negative over a range of $m_{val} \lessgtr m_{sea}$. Nevertheless, to
		understand the sign of scalar correlators in various phases, we
		certainly need corresponding mixed action ChiPT. 
\section{$\Delta_{\rm mix}$\label{sec_deltamix}}
\begin{figure}
\begin{center}
{ (a)}\includegraphics[width=8cm,clip]{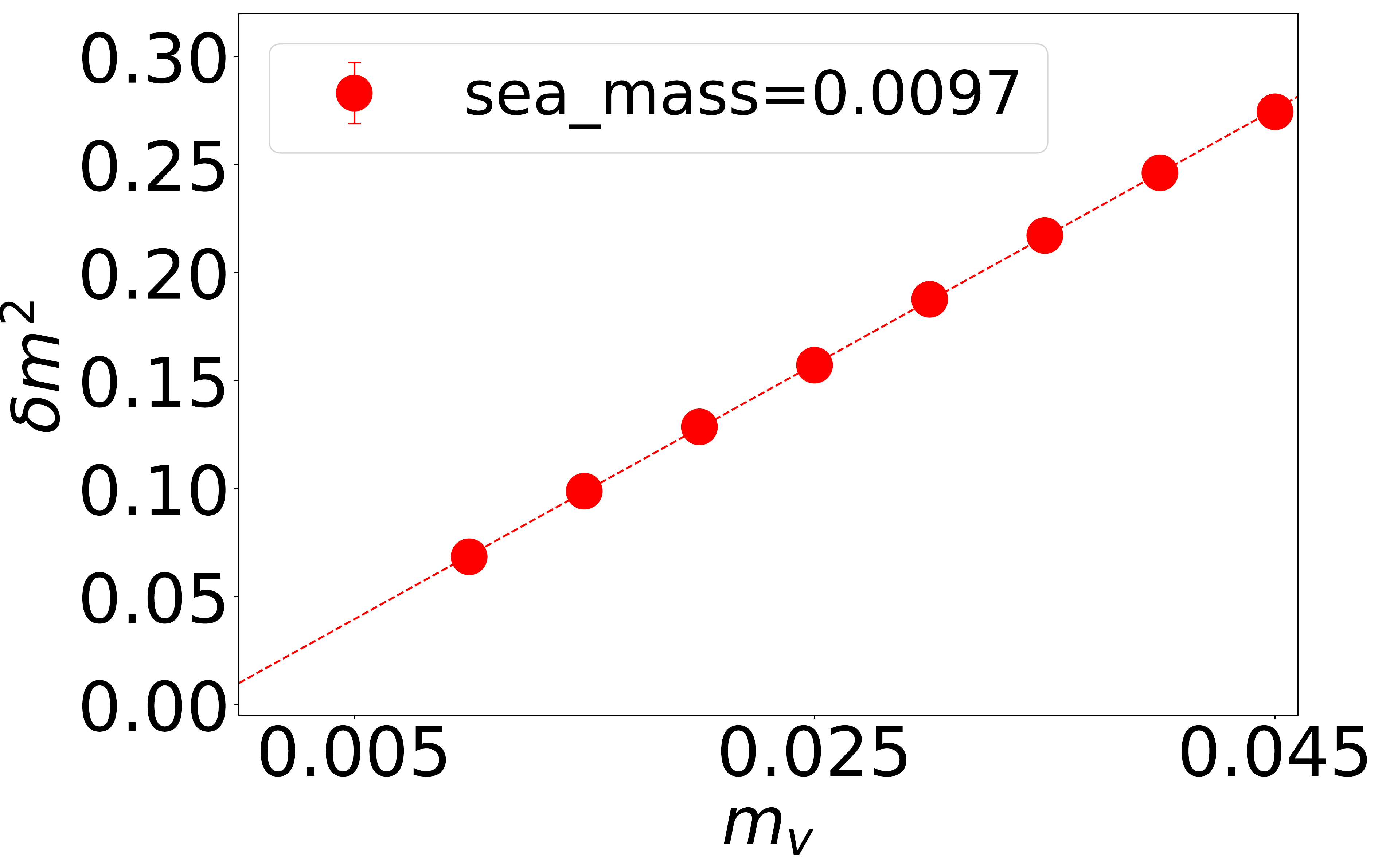}\\
{(b)} \includegraphics[width=8cm,clip]{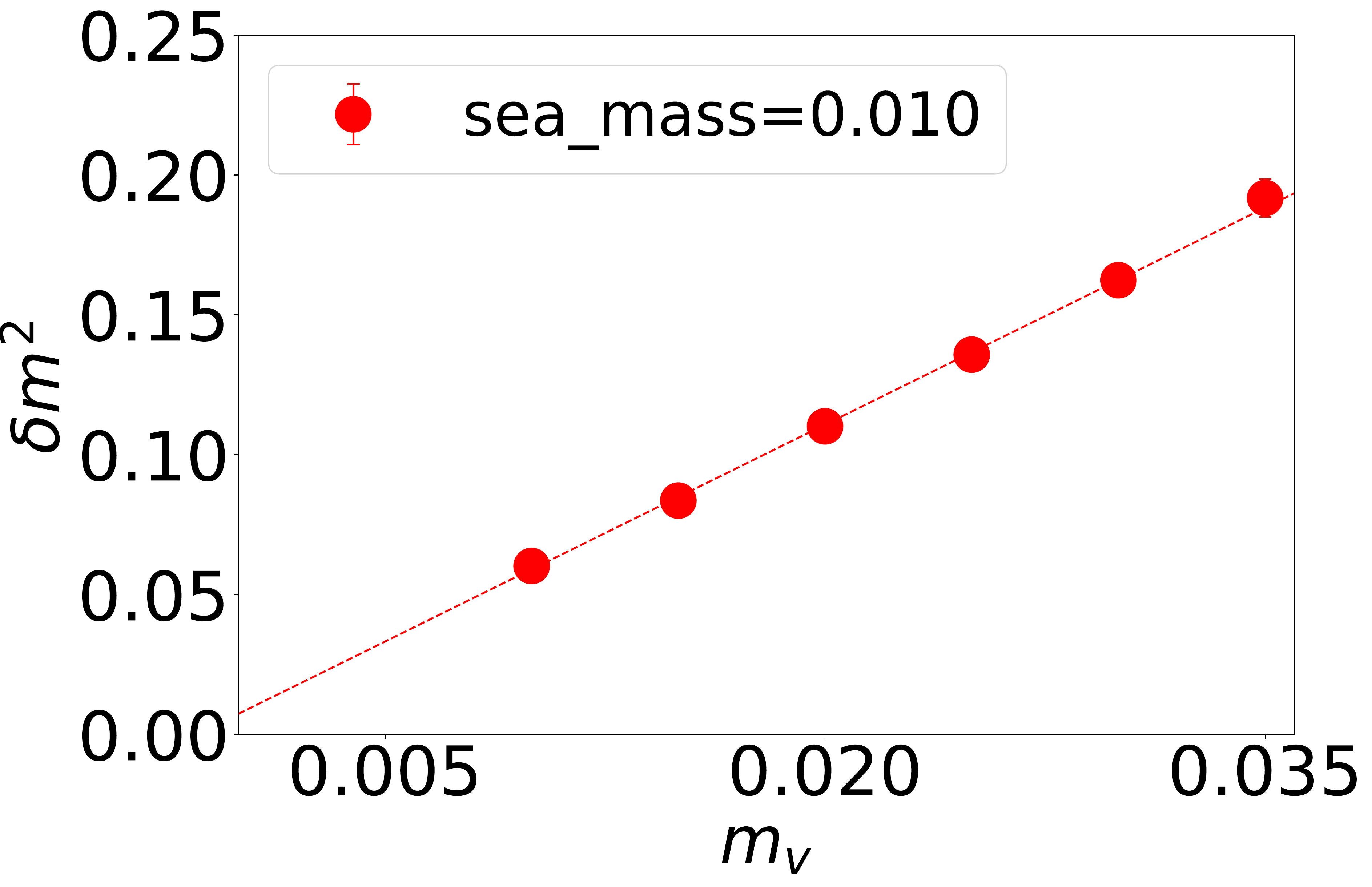}
\caption{$\delta m^2=m_{vs}^2-\frac{1}{2}m_{ss}^2$  plotted as a
function of $m_v$ (a) $16^3\times 48$ lattice with $m_s=0.097$,
(b) $20^3\times 64$ lattice with $m_s=0.01$. The intercept on the
$\delta m^2$ axis gives the value of $\tilde{\Delta}_{mix}$.}
\label{delta_mix}
\end{center}
\end{figure}

In mixed action calculations, we can have three different types of
mesons: mesons composed of ($i$) two valence quarks, ($ii$) two sea
quarks and ($iii$) a mix of one valence and one sea quark. Each of these
undergo lattice spacing dependent mass renormalization. The mixed action
$\chi$PT in leading order has an $\mathcal{O}(a^2)$ dependent low energy 
constant $\Delta_{\rm mix}$. The degree of unitarity violation at finite
lattice spacing depends on the size of $\Delta_{\rm mix}$. In the leading 
order, the psuedoscalar meson masses for BC valence and Asqtad sea are
given by
\be
m_{v_1 v_2}^2 &=& B_v(m_{v_1}+m_{v_2}) \label{mes_vv} \\
m_{s_1s_2,t}^2&=& B_0(m_{s_1}+ m_{s_2}) + a^2\Delta_t \label{mes_ss} \\
m_{vs}^2 &=& B_v m_v + B_0 m_s + a^2\tilde{\Delta}_{\rm mix},
\label{mes_vs}
\ee
where $m_{v_1v_2}$ ($m_{s_1s_2}$) is the pseudoscalar meson mass made
up of valance (sea) quark and antiquark while $m_{vs}$ is the mass of
the mixed meson composed of valence and sea quarks. The $a^2\Delta_t$
are the taste splittings of the {\em staggered} pions, where $t = A,\,
T,\, V, \, I$ and $a^2 \Delta_5 =0$ \cite{aubin1} and $a^2 \tilde{
\Delta}_{\rm mix} = a^2\Delta_{\rm mix} + a^2\Delta_{\rm mix}^\prime$
where \cite{golterman},
\be
a^2 \Delta_{\rm mix}^\prime = \frac{1}{8} a^2 \Delta_A + \frac{3}{16}
a^2 \Delta_T + \frac{1}{8} a^2 \Delta_V + \frac{1}{32} a^2 \Delta_I.
\label{deltamixp}
\ee
The different renormalizations of the quark masses in different
actions are absorbed in the coefficients $B_0$ and $B_v$. The
$\tilde{\Delta}_{\rm mix}$ can be extracted from the meson spectrum
data from either of the following ways,
\be
&& m_{vs}^2-\frac{1}{2}m_{vv}^2= B_o m_s+
a^2\tilde{\Delta}_{\rm mix}  \;\;\; {\rm or} \label{dmix1} \\
\delta m^2 & \equiv & m_{vs}^2-\frac{1}{2}m_{ss,5}^2 =  B_v m_v+a^2
\tilde{\Delta}_{\rm mix}. \label{m_diff}
\ee
It is convenient to work with the form in \autoref{m_diff} as there are
more $m_v$ available (than $m_s$) for a good linear fit. The Fig.
\ref{delta_mix} shows the result of variation of $\delta m^2$ with
$m_v$ and the linear extrapolation of $\delta m^2$ in the bare valence
mass $m_v$ gives the $a^2 \tilde{\Delta}_{\rm mix}$ as the $y$-intercept.
To determine the actual size of unitarity violation due to differences
in valence and sea quark discretization, one needs to subtract the taste
splitting dependent terms in \autoref{deltamixp} \cite{deltaprime,DWF1}.

\vskip 0.1in
\begin{table}[hpt] 
\centering \begin{tabular}{|c|c|c|}
\hline Mixed action &  $a$ fm & $\tilde{\Delta}_{mix}$ GeV$^4$  \\
\hline Overlap on Clover\cite{Clover} & 0.09 & 0.55(23)\\
DWF on staggered\cite{DWF2} & 0.125 & 0.249(6)\\
DWF on staggered\cite{DWF1} & 0.12 & 0.211(16)\\
DWF on staggered\cite{DWF1} & 0.09 & 0.173(39)\\
Overlap on DWF\cite{Overlap2} & 0.114 & 0.030(6)\\
Overlap on DWF\cite{Overlap2} & 0.085 & 0.033(12)\\  
Overlap on HISQ \cite{hisq} & 0.12 & 0.112(11)\\
BC on staggered[this work] & 0.15 & 0.03(1)\\
BC on staggered[this work] & 0.13 & 0.03(8)\\
\hline 
\end{tabular}
\vskip 0.1in
\caption{$\tilde{\Delta}_{\rm mix}$ for different mixed actions with
$300$ MeV pion mass.}
\label{delta_comp}
\end{table}

The values of $\tilde{\Delta}_{\rm mix}$ have been tabulated and compared
with other works in Table \ref{delta_comp}. For the form of $a^2 \Delta_{
\rm mix}^\prime$ for Wilson sea fermions see \cite{deltaprime}. The value
of $\tilde{\Delta}_{\rm mix}$ obtained in this work is rather encouraging.
Although our value appears smaller than the studies with domain wall
fermion on staggered sea \cite{DWF1,DWF2}, a direct comparison with it
is difficult because of our coarser lattices, less statistics, fitting
strategy and no systematic error estimation. Note that we feature the
results for $\tilde{\Delta}_{\rm mix}$ that includes the $\Delta_{\rm
mix}^\prime$, which in our case contains contribution from the taste
splitting. Still, the lattice artifact $\tilde{\Delta}_{\rm mix}$,
measuring the size of unitarity violation in our mixed action lattice
simulations with Borici-Creutz fermions on staggered sea, is of the
same order of magnitude as the rest.

\section{Summary and Conclusions}
Mixed action lattice QCD is commonly employed to compute hadronic
observables as it allows use of numerically cheaper lattice fermions
in the sea sector while using chirally improved, and possibly expensive,
valence fermions. Such mixed action approach is also rather expedient
when a new fermion discretization, Borici-Creutz fermions in the present
case, is tried for lattice calculations. In this paper, we presented
the first results for light hadron mass spectrum and mixed action
parameter $\Delta_{\rm mix}$ using Borici-Creutz valence fermion on
staggered sea.

As a first step, we nonperturbatively tuned the two counterterm
coefficients $\tilde{c}_3$ and $c_4$ to restore the parity and time
symmetry, which are otherwise broken by Borici-Creutz action. We found
that the tuned values of $\tilde{c}_3 = -0.50$, $c_4 = 0.005$ are
significantly different from the perturbative estimates (given in
Table \ref{pert}). We observed that the tuning of $\tilde{c}_3$ and
$c_4$ can proceed independently of each
other. The $\tilde{c}_3$ is tuned by minimizing the value of parity
condensate $\vert \langle i\,\bar{\psi} \gamma_5 \tau^3 \psi \rangle
\vert$. The variation of parity condensate with $\tilde{c}_3$ is found
to be insensitive to variation of $c_4$ over the range [0.005,\, 0.1]
and three different lattice spacings 0.15\,fm, 0.13\,fm, 0.09\,fm.

In a (discretized) Lorentz invariant theory, the forward and backward
propagating pions are the same and have the same masses. But with the
Borici-Creutz action this has to be achieved by tuning the counterterm
coefficients by minimizing the mass differences of the forward and
backward propagating pions. We observed that the forward-backward
symmetry, {\em i.e.} the time symmetry, is attained entirely by tuning
$c_4$, independent of $\tilde{c}_3$ over the range [-1.8,\, -0.4].
The term containing $c_4$ is a kinetic `like' term and, therefore,
expected to influence the correlation of operators separated over
time. This might be the reason we see pion correlation function
propagating differently in opposite time direction. We must clarify
that this differently propagating pion in time is not the same as
opposite parity baryon and antibaryon propagating forward and backward
in time.

The tuned values of $\tilde{c}_3$ and $c_4$ seem to have very small
dependence on the lattice spacings. This is perhaps expected since
the observables used for tuning are not known to have significant
cut-off effect. We don't expect these tuned values to change much
with different $m_l/m_s$ and physical volumes but certainly will
depend on the variant of sea fermions.

Once the counterterms are nonperturbatively tuned,
we studied the variation of pion mass $m_\pi$ with the bare
valence quark masses $m$ in the range [0.007 -- 0.5]. The plot
$m_\pi^2$ versus $m$ in Fig. \ref{mpi_vs_mq} shows almost
linear behavior in the entire range of quark mass except near
small masses. The partially quenched chiral log is most visible
in the plot $m_\pi^2/m$ versus $m$, the behavior of which
can be described by PQ$\chi$PT formula \autoref{chiral_fit}.
Since we did not use simultaneous {\em i.e.} global fit over
all the ensembles, we cannot comment much on the parameters
$C_i$'s of \autoref{chiral_fit}. They are used only for
fitting purpose and hence not quoted. The chiral log plot has
been used in \cite{Overlap2} to determine the range of quark
masses where LO MA$\chi$PT relation(s) can be used. In our case,
we used both $m_\pi^2$ vs $m$ and $m_\pi^2/m$ vs $m$
plots to determine the valence quark mass range. It is to be
noted that the mass tuning in mixed action simulation is not
unique and for any choice the mixed action will violate unitarity
at finite lattice spacing and all partial quenching pathologies
will show up. For instance, this tuning can also be done by
matching the pion mass in valence sector to the pion mass in the
sea sector. Hence, in this work we chose quark masses that are
just outside the region of chiral log, ensure that $m_\pi \,L
\geq 4$ and $m_{\rm val} \gtrsim m_{\rm sea}$.

The scalar correlators, plotted in Figs. \ref{scalar_corr} shows
the unitarity violating effects in the mixed action theory. However,
unlike other mixed action scalar correlator plots, there is an extra
complication in the behavior of the scalar correlators -- it is the
role of the counter-term $\tilde{c}_3$. The term $\tilde{c}_3\,\bar{
\psi} \Gamma \psi$, is like a mass 'like' term and, as a result,
we do not see an `absolute' clear negative values in the scalar
correlators. In the plot for $\tilde{c}_3 = -0.5$, the scalar
correlator for $m = 0.0055 \; ( < m_l)$ has negative mean values
but the error bars are large implying correlators varying over a
range of positive and negative values. But, when $m = 0.02 \;
(> m_l)$, the correlator points are distinctly in the positive
region. From the other plot with $m = 0.0055$ and $\tilde{c}_3
= -1.6211$, where parity is only partially restored, we find that
$\tilde{c}_3$ is driving the scalar correlators towards positive
value regardless the fact $m < m_l$. An MA$\chi$PT for Borici-Creutz
valence fermion on staggered sea, which is presently not available,
can only tell us exactly how this is achieved.

Finally, we want to determine the size of unitarity violation by
measuring the mixed valence-sea meson mass splitting $\Delta_{\rm
mix}$. This is an important lattice artifact to determine for any
mixed action calculations. In this paper we actually measured the
$\tilde{\Delta}_{\rm mix}$, which contains an additional lattice
spacing dependent term $\Delta_{\rm mix}^\prime$ containing the
contributions of the taste splittings. For pion mass about 300 MeV,
we obtained $\tilde{\Delta}_{\rm mix} = 0.03(1)$ (GeV)$^4$ at $a
= 0.15$ fm and 0.03(8) (GeV)$^4$ at $a = 0.13$ fm, {\em i.e.}
they are same within the error. This result is at par (same order
of magnitude) with the other mixed action studies as tabulated in
Table \ref{delta_comp}. However, the fits that are performed here
to arrive at these results are all uncorrelated and no global
fitting (involving all the ensemble data) are performed. Thus the
mean value and error bar of $\tilde{\Delta}_{\rm mix}$ can change
depending on more sophisticated fitting procedure. But we expect
it not to be substantial. Based on this observation, we can
certainly claim that this is an encouraging first result for
Borici-Creutz fermion when compared to other mixed action studies.
It is worth following up the study of Borici-Creutz fermion more
rigorously and attempt a fully unquenched lattice QCD investigation.

{\bf Acknowledgements: }
The numerical jobs have been run on the HPC at IIT Kanpur funded by DST and 
IIT Kanpur and on computer 
facility at NISER, Bhubaneswar. One of the authors (SB) thanks DST-SERB for 
providing fund
(project number SR/S2/HEP-0025/2010) for computers in which a
part of this work is carried out and JG thanks  Rudina 
Zeqirllari for useful communications.


\end{document}